\documentclass[aps,pra,twocolumn,superscriptaddress,showpacs,showkeys]{revtex4-2}

\usepackage{amsmath}
\usepackage{amssymb}
\usepackage{braket}
\usepackage{color}
\usepackage{dsfont}
\usepackage{graphics}
\usepackage{graphicx}
\usepackage{slashed}
\usepackage[hidelinks]{hyperref}
\usepackage{eufrak}
\renewcommand{\vec}[1]{\boldsymbol{#1}}

\begin{document}

\preprint{APS/123-QED}

\title{Two-dimensional simulation of the spin-flip in the Kapitza-Dirac effect}
\author{Ping Ge}
\author{Sven Ahrens}
\thanks{ahrens@shnu.edu.cn}%
\author{Baifei Shen}%
\email{bfshen@shnu.edu.cn}
\affiliation{Shanghai Normal University, Shanghai 200234, China}%

\begin{abstract}
Many calculations in strong field quantum field theory are carried out by using a simple field geometry, often neglecting the spacial field envelope. In this article, we simulate the electron diffraction quantum dynamics of the Kapitza-Dirac effect in a Gaussian beam standing light wave. The two-dimensional simulation is computed in a relativistic framework, by solving the Dirac equation with the fast Fourier transform split operator method. Except the numerical propagation method, our results are obtained without applying approximations and demonstrate that a spin-flip in the Kapitza-Dirac effect is possible. We further discuss properties, such as the validity of a plane wave approach for the theoretical description, the influence of the longitudinal polarization component due to laser beam focusing and higher order diffraction peaks in Kapitza-Dirac scattering.
\end{abstract}

\maketitle


\section{\label{sec:level1} Introduction}

In present day strong laser fields, it is possible to facilitate spin effects for free electrons \cite{Chen_Keitel_2019_polarized_positrons,  Li_Keitel_Li_2019_single_shot_polarization, Li_Chen_2020_polarized_positron_electron,   Wen_Tamburini_Keitel_2019_polarized_kilo_ampere_beams,  Del_Seipt_Blackburn_Kirk_2017_electron_spin_polarization,   Karlovets_2011_radiative_polarization,   van_Kruining_Mackenroth_2019_magnetic_field_polarization,Keitel_2023_one_loop_polarization}. One particular variation of spin-laser interaction of electrons is the Kapitza-Dirac effect \cite{kapitza_dirac_1933_proposal,Freimund_Batelaan_2001_KDE_first}, for which spin effects are predicted \cite{ahrens_bauke_2012_spin-kde,ahrens_bauke_2013_relativistic_KDE,dellweg_awwad_mueller_2016_spin-dynamics_bichromatic_laser_fields,erhard_bauke_2015_spin}, in a scenario which is similar to Bragg scattering \cite{Bragg_1913_Bragg_scattering,batelaan_2000_KDE_first,batelaan_2007_RMP_KDE}. The setup of Kapitza-Dirac scattering, in which an electron traverses a standing light wave, formed by two counterpropagating beams, can be tailored to be sensitive to the spin polarization of the incoming electron \cite{dellweg_mueller_2016_interferometric_spin-polarizer,ahrens_2017_spin_filter,dellweg_mueller_extended_KDE_calculations,ahrens_2020_two_photon_bragg_scattering,McGregor_Batelaan_2015_two_color_spin,ebadati_2018_four_photon_KDE,ebadati_2019_n_photon_KDE}. Therewith the effect is allowing for a laser based Stern-Gerlach type spin observation \cite{Stern_Gerlach_1922_1,Stern_Gerlach_1922_2,Stern_Gerlach_1922_3}, in form of an induced Compton scattering process \cite{Compton_1923_compton_scattering,ahrens_2020_two_photon_bragg_scattering}, being a fundamental photon-only interaction. Experiments in the Bragg regime exist \cite{Freimund_Batelaan_2002_KDE_detection_PRL}, even with observing the cancellation of the interaction at parameters, where spin effects are expected \cite{Axelrod_2020_Kapitza_Dirac_cancellation_observation}.

Most theoretical descriptions of the Kapitza-Dirac effect implement the standing wave potential of the external field by two counterpropagating plane waves, where the field's width and longitudinal polarization component are neglected. Since Gaussian beam solutions \cite{Dickson_1970_gaussian_beam} can be considered to be more realistic than a plane wave approach, we were investigating the Gaussian beam influence in a recent study on spin-dynamics in Kapitza-Dirac scattering  \cite{ahrens_guan_2022_beam_focus_longitudinal}. In order to solve the problem analytically, rough approximations were imposed on the plane wave approach. One of the approximations was the assumption of a discrete set of plane wave superpositions, for solving the relativistic equations of motion of the Dirac equation in the perturbative approach \cite{ahrens_guan_2022_beam_focus_longitudinal}.  Naturally, the question arises, whether the approximations of the standing wave vector potential within a perturbative solution technique are sufficiently accurate. In this article, we solve the quantum dynamics of the electron wave function on a two-dimensional grid, by using a Fast-Fourier-transform (FFT) split-operator method \cite{Grobe_1999_FFT_split_operator_method,bauke_2011_GPU_acceleration_FFT_split_operator}. Within this method, the Gaussian beam potential can be implemented exactly, such that no approximations need to be applied to the external field. This work is thus a demonstration of spin-flip dynamics of an electron in the Kapitza-Dirac effect on the basis of a relativistic, two-dimensional simulation, in which the Dirac equation is evolved numerically. We are further able to study the role of the longitudinal polarization component on the quantum dynamics in our work, as well as the validity of a plane wave approach for describing the effect.

Our article is organized as follows. In section \ref{sec:setup_of_investigation}, we discuss the simulation setup, by introducing the Gaussian laser beam (section \ref{sec:gaussian_beam}), the relativistic quantum description (section \ref{sec:relativistic_quantum_theory}) and the initial condition of the electron quantum state (section \ref{sec:initial_quantum_state}). We also mention details about simulation parameter configuration, as well as the numerical procedure of the Q-Wave library in section \ref{sec:simulation_parameters}. We then present the simulation results in section \ref{sec:simulation_results}. The results include the demonstration of electron diffraction dynamics in the Kapitza-Dirac effect (section \ref{sec:electron_motion}), with displaying the spin properties of the quantum dynamics in section \ref{sec:spin_dynamics}. In section \ref{sec:further_analysis} we further investigate physical properties of the Kapitza-Dirac effect, such as the validity of the commonly used plane approximation for describing the Kapitza-Dirac effect (section \ref{sec:plane_wave_validity}), frequency and beam focus scaling of the spin-dynamics (section \ref{sec:scaling_effects}) and the influence of the longitudinal beam polarization component of focused beams on the effect (section \ref{sec:longitudinal_component}). In section \ref{sec:diffraction_regime} we also have a look on the emergence of higher order diffraction peaks in Kapitza-Dirac scattering. Finally, we summarize our investigation and give an outlook on possible, related topics in section \ref{sec:conclusions}.

\section{Setup of our investigation\label{sec:setup_of_investigation}}
For our computer simulation we make use of the Q-Wave utility \cite{bauke_2011_GPU_acceleration_FFT_split_operator}. Q-Wave is an advanced computer code, available as C++ library, which implements the FFT split-operator method, among other numerical algorithms \cite{Beerwerth_2015_Krylov_subspace_methods}. It provides the building blocks for numerically propagating wave functions in time. In the following, we describe the physical setup which we investigate by using Q-Wave. Regarding the units in our article, we write $m$ for the electron rest mass, $c$ for the vacuum speed of light, $\hbar$ for the reduced Planck constant and $q$ for the elementary charge in a Gaussian unit system.

\subsection{Gaussian beam configuration\label{sec:gaussian_beam}}

We first describe the vector potential of our simulation. A Gaussian beam shaped standing light wave can be formed from two Gaussian beams \cite{Quesnel_1998_gaussian_beam_coulomb_gauge}, where reference \cite{Quesnel_1998_gaussian_beam_coulomb_gauge} builds on a solution based on an angular spectrum representation of plane waves. The laser beam is propagating along the $x$-axis, in our two-dimensional simulation, where the simulation area is aligned in the $x$-$y$ plane. For the geometry in this article, the Gaussian beam is denoted as
\begin{subequations} \label{Equation_of_Gaussian_beam}
\begin{equation}
A_{x,d}=-2dA_{0}\frac{w_{0}}{w}\epsilon\frac{y}{w}\exp \left(-\frac{r^{2}}{w^{2}} \right)\cos \left(\phi^{(1)}_{G,d}\right)\label{Equation_of_Gaussian_beam_longitudinal}
\end{equation}
for the longitudinal polarization component and
\begin{equation}
A_{y,d}=-A_{0}\frac{w_{0}}{w}\exp \left(-\frac{r^{2}}{w^{2}} \right)\sin{\left(\phi_{G,d} \right)}\label{Equation_of_Gaussian_beam_transverse}
\end{equation}
\end{subequations}
for the transverse polarization component of the vector potential in Coulomb gauge \footnote{Further details about adjusting the fields in \cite{Quesnel_1998_gaussian_beam_coulomb_gauge} can be found in the appendix of reference \cite{ahrens_guan_2022_beam_focus_longitudinal}.}. The potentials in Eq. \eqref{Equation_of_Gaussian_beam} further contain the phases
\begin{subequations} \label{Phase_of_Gaussian_beam}
\begin{align}
\phi_{G,d}&=\omega t-dk_{L}x+\tan^{-1} \left(\frac{dx}{x_{R}} \right)-\frac{dxr^{2}}{x_{R}w^{2}}\\
\phi_{G,d}^{(1)}&=\phi_{G,d}+\tan^{-1} \left(\frac{dx}{x_{R}} \right)
\end{align}
\end{subequations}
and the symbol $w$ represents the $x$-dependent beam waist
\begin{equation}
w(x)=w_{0}\sqrt{1+\frac{x^{2}}{x^{2}_{R}}} \,.
\end{equation}
The Gaussian beam oscillates with frequency $\omega$, with corresponding wave number $k_L=\omega/c$ and wavelength $\lambda=2\pi /k_{L}$. Further, reference \cite{Quesnel_1998_gaussian_beam_coulomb_gauge} introduces the beam focus as $\omega_{0}=1/(k_{L}\epsilon)$, with Rayleigh length of the Gaussian beam  $x_{R}=k_{L}\omega_{0}^{2}/2$.

\begin{figure}
\includegraphics[width=0.48 \textwidth]{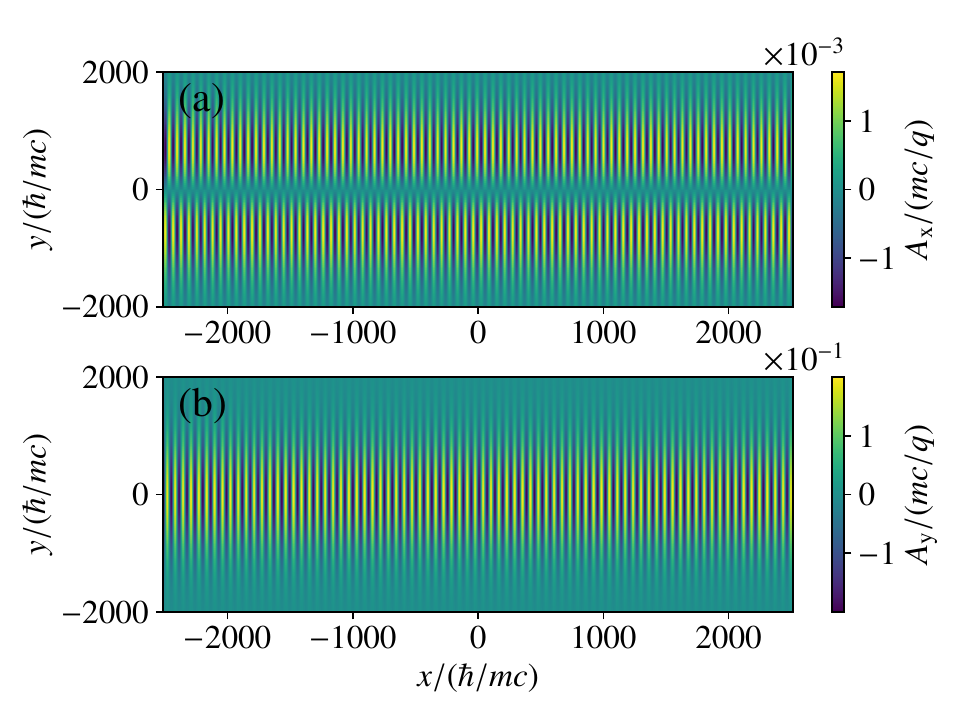}
\caption{\label{fig:Simulation_of_Gaussian_beam}Vector potential of a Gaussian beam standing wave in our simulation, according to Eq. \eqref{Equation_of_vector_potential}. The longitudinal \eqref{Equation_of_Gaussian_beam_longitudinal} and transverse \eqref{Equation_of_Gaussian_beam_transverse} polarization components of the laser beam are displayed at the upper and lower panel, respectively. We use the laser parameters $A_0 = 0.1\,mc/q$ for the field amplitude, $k_L=0.1\,mc/\hbar$ for the wave number and $\epsilon=0.02$ for the beam divergence, in the Gaussian beam, displayed after at a quarter laser period $t=\omega/(2\pi)$. In our two-dimensional simulation area, the electron passes the Gaussian beam from bottom to top, along the $y$-direction.}
\end{figure}

The index $d$ in Eqs. \eqref{Equation_of_Gaussian_beam} parameterizes the propagation direction of the laser beam, where the two possible directions  $d\,\in\,\{-1,1\}$ correspond to the left or right moving direction, respectively. The standing wave vector potential in the Kapitza-Dirac effect can be formed from the two counterpropagating beams by the superposition
\begin{equation}
 \vec A = \sum_d \left( A_{x,d} \vec e_x + A_{y,d} \vec e_y \right)\,.\label{Equation_of_vector_potential}
\end{equation}
We display the field $\vec A$ in Fig. \ref{fig:Simulation_of_Gaussian_beam} as it appears after a quarter laser period $t=\omega/(2\pi)$ for the parameters of our showcase simulation in section \ref{sec:simulation_parameters}. In contrast to previous theoretical investigations, transverse and longitudinal polarization are both computed without applying approximations here, with a finite beam width and a longitudinal polarization component.

\subsection{Relativistic quantum theory\label{sec:relativistic_quantum_theory}}

Since the laser field in our simulation is strong and the initial electron momentum of electron is $1 mc$, we use a relativistic spin 1/2 quantum theory for the description of our simulation, which is given by the electromagnetically coupled Hamiltonian of the Dirac equation
\begin{equation} \label{Hamiltonian operator}
H = c \left( \vec{p} - \frac{q}{c} \vec{A} \right) \cdot \vec{\alpha}+q\phi+\beta mc^{2} \,.
\end{equation}
The gauge potential $\vec{A}$ has been introduced in subsection A, where set the scalar potential to zero $\phi=0$ in our code. The objects $\vec \alpha$ and $\beta$ are the $4\times4$ Dirac matrices in standard representation (also called Dirac representation). We write the energy eigenvalue relations in momentum space as
\begin{equation}\label{Eigenfunction_of_electron}
H \psi^{s}({\vec{p}})= E(\vec p) \psi^{s} (\vec{p}) \,,
\end{equation} 
with the positive plane wave solutions of the Dirac equation
\begin{equation}\label{Solution_of_Dirac_equation}
\psi^{s} (\vec{p}) =u^{s}(\vec{p}) e^{i\vec{r}\cdot\vec{p}/ \hbar} \,,
\end{equation} 
where the we denote the bi-spinors $u^{s}(\vec p)$ as
\begin{equation} \label{Spinors_of_Dirac_equation}
u^{s}(\vec{p})=\sqrt{\frac{E(\vec p)+mc^{2}}{2mc^{2}}} \begin{pmatrix}
\chi^{s} \\ \frac{c \vec{\sigma} \cdot \vec{p}}{E(\vec p)+mc^{2}}\chi^{s}
\end{pmatrix} \,.
\end{equation}
In Eqs. \eqref{Eigenfunction_of_electron}-\eqref{Spinors_of_Dirac_equation}, the parameter $s\in \{+,- \}$ is indexing the state of the electron spin with the $x$-polarized basis
\begin{equation}
\chi^+=
\begin{pmatrix}
 1 \\ 1
\end{pmatrix}\,,\qquad
\chi^-=
\begin{pmatrix}
 1 \\ -1
\end{pmatrix}\,.
\end{equation}
We also write
\begin{equation}
E(\vec p)=\sqrt{m^{2}c^{4}+c^{2} \vec p^{2}}\label{eq:relativistic_energy_momentum_relation}
\end{equation}
for the relativistic energy, $\vec{p}=p_{x} \cdot \vec{e} _{x}+ p_{y} \cdot \vec{e}_{y}$ for the momentum vector and $\vec{\sigma}$ for the vector of Pauli matrices.

\subsection{The initial electron quantum state\label{sec:initial_quantum_state}}

According to the Q-Wave simulation package \cite{bauke_2011_GPU_acceleration_FFT_split_operator}, the wave packet of the electron is initialized as a Gaussian wave packet, in our two-dimensional simulation, with the density distribution
\begin{equation} \label{Density_distribution_of_Gaussian_wave_packet}
\rho(\vec p) = \frac{1}{\sqrt{2 \pi} \sigma_p} \exp \left[- \left( \frac{\vec{p}-\vec{p}_{0}}{2\sigma_{p}}\right)^{2} -i\frac{\vec r_0 \cdot \vec p }{\hbar}  \right]
\end{equation}
in momentum space. The Gaussian distribution is centered at momentum $\vec p_0$, with wave packet size parameter $\sigma_p$. The second term in the exponential implies the particle's position at $\vec r_0$. The wave function in momentum space is set up as
\begin{equation}
\varphi(\vec{p},0)=u^+(\vec{p})\rho(\vec p) \,,
\end{equation}
on the basis of the distribution \eqref{Density_distribution_of_Gaussian_wave_packet}. In position space, the wave function $\Psi$ is then implied by the two-dimensional Fourier transformations
\begin{subequations}\label{eq:fourier_transformation}%
\begin{align}%
\varphi(\vec{p},t)&=\frac{1}{2\pi \hbar} \int \Psi(\vec{r},t) \exp \left(-\frac{i \vec{r} \cdot \vec{p}}{\hbar} \right) d^{2}r\label{eq:fourier_transformation_of_position_space}\\
\Psi(\vec{r},t)&=\frac{1}{2\pi \hbar} \int \varphi(\vec{p},t)   \exp \left(\phantom{-}\frac{i \vec{r} \cdot \vec{p}}{\hbar} \right) d^{2}p\,.\label{eq:fourier_transformation_of_momentum_space}
\end{align}%
\end{subequations}%


\subsection{Numerical propagation and simulation parameters\label{sec:simulation_parameters}}

The Q-Wave library provides numerical algorithms for solving the time-evolution of quantum wave function in multiple time steps. We make use of the Fast Fourier split operator method \cite{bauke_2011_GPU_acceleration_FFT_split_operator}, for which the a time step with time stepping $\Delta t$ can be denoted as a mapping of the wave function $\Psi{(r,t)}$ to the wave function $\Psi{(r,t + \Delta t)}$ at a later point in time by
\begin{equation}
\Psi(\vec{r},t+\Delta t)=\vec{U}(t+\Delta t,t)\Psi(\vec{r},t) \,.
\end{equation}

\begin{figure}
\includegraphics[width=0.48 \textwidth]{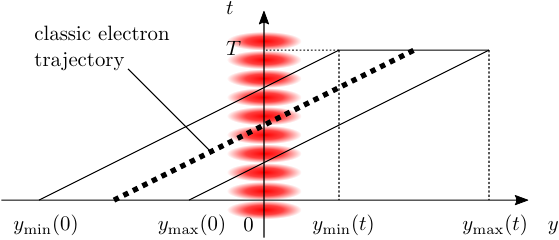}
\caption{\label{KDE_simulation_process} Illustration of the simulation box limits and electron location along the  $y$-direction. The red, oval shaped shades symbolize the laser beam position over time, located at $y=0$. Along the $y$-direction, the simulation box size equals $40\lambda$ with initial minimum and maximum positions $y_{\textrm{min}}(0)=-160\lambda$ and $y_{\textrm{max}}(0)=-120\lambda$, respectively, at time $t=0$. The electron is initially placed in the simulation box center, at $y=-140\lambda$ and moves approximately along the classical electron trajectory $y_\textrm{CET}(t)$, as given in Eq. \eqref{eq:classical_electron_trajectory_y_component}, which we indicate by the thick, dotted line. 
In the supplemental material, we provide an animation of the vector potential as in Fig. \ref{fig:Simulation_of_Gaussian_beam}, within the moving bounds of the parameterized $y$-axis, as sketched here.}
\end{figure}

In the following we will introduce specific values of parameters, which are set in the simulation. We carry out our simulation on a grid with $2048\times 128$ grid points, with simulation area width $80\lambda$ and height $40\lambda$, in the $x$ and $y$ direction, respectively. Along the $x$-axis we set the minimum and maximum simulation box limits $x_\textrm{min}=-40\lambda$ and $x_\textrm{max}=40\lambda$. For the $y$-axis, we require the electron wave packet to be centered in our simulation area, initially and during the simulation, as sketched in Fig. \ref{KDE_simulation_process}. We choose the initial simulation box limits as $y_\textrm{min}(0)=-160\lambda$ and $y_\textrm{max}(0)=-120\lambda$, corresponding to a distance of approximately 15 half beam waists $w_0$ away from the laser beam center.

The electron's initial position along the $y$-direction is in the simulation box center at $y=-140 \lambda$. Regarding the electron's momentum, we set the momentum $p_x = - \hbar k_L$ along $x$-axis, to meet the Bragg condition for the two-photon Kapitza-Dirac effect \cite{ahrens_bauke_2012_spin-kde,ahrens_bauke_2013_relativistic_KDE,ahrens_2012_phdthesis_KDE}. The $y$-component of the electron momentum is implied by the requirement for spin effects in the Kapitza-Dirac effect \cite{ahrens_bauke_2013_relativistic_KDE,ahrens_2017_spin_non_conservation, ahrens_2020_two_photon_bragg_scattering} to be $p_y = 1 mc$.
The momentum parameter $\vec p_0$ for the initial electron state in Eq. \eqref{Density_distribution_of_Gaussian_wave_packet} is therefore assuming the value

\begin{equation}\label{eq:electron_momentum}
 \vec p_0 =
\begin{pmatrix}
 -\hbar k_L \\ m c
\end{pmatrix}\,,
\end{equation}
with inclination angle of the Bragg condition
\begin{equation}
  \vartheta = \arctan(|p_x|/|p_y|)\,.
\end{equation}
Requiring that the electron needs to move through the coordinate origin, this also implies that the initial electron position along the laser beam propagation direction has to be $x=140\lambda_C$, such that the initial position vector in Eq. \eqref{Density_distribution_of_Gaussian_wave_packet} reads
\begin{equation}
\vec r_0 =
\begin{pmatrix}
 140 \lambda_C \\ -140 \lambda
\end{pmatrix}\,,\label{eq:electron_position}
\end{equation}
where $\lambda_C=h/(m c)$ is the Compton wavelength with the Planck constant $h=2\pi \hbar$.

We set the momentum spread of the electron to $\sigma_p = \hbar k_L/200$, which corresponds to an electron wave function extension on the order of 100 laser wave lengths. Concerning the simulation time, we mention that the significant $y$-component of the electron momentum \eqref{eq:electron_momentum} implies the approximate classical electron velocity $v_y=c/\sqrt{2}$, with the corresponding $y$-component of the classical electron trajectory
\begin{equation}
 y_\textrm{CET}(t)=-140 \lambda + \frac{1}{\sqrt{2}} ct\,. \label{eq:classical_electron_trajectory_y_component}
\end{equation}
Eq. \eqref{eq:classical_electron_trajectory_y_component} implies the traveling time $\tilde T=280\lambda\sqrt{2}/c=2.5\cdot10^{4}\hbar/(mc^2)$, 
if we require the electron to move up to the $y$-axis position $y=140 \lambda$. Further, we choose the time stepping $\Delta t=0.05\,\hbar/{m c^2}$, for resolving the oscillation of the mass term $\beta m c^2$ in the Dirac equation.

For the Gaussian beams of our standing light wave, we set the parameters $\omega=0.1\,\hbar/mc$ for the laser angular frequency, $A_{0}=0.1\,mc/q$ for the field amplitude and $\epsilon=0.02$ for the beam divergence, in Eqs \eqref{Equation_of_Gaussian_beam} and \eqref{Phase_of_Gaussian_beam}. We remark, that for technical reasons, we introduce a shift between the kinetic and canonical momentum of the wave packet by employing a gauge with constant vector potential $A_m = 1.0\, mc/q$ in the $y$-axis of Eq. \eqref{Equation_of_vector_potential} in our numeric implementation. We further point out that the computation of the quantum state time evolution is numerically implemented in a $z$-polarized spinor basis, where the $x$-polarized description with the spinors \eqref{Spinors_of_Dirac_equation} is obtained from a superposition of the $z$-polarized basis, as the Dirac equation is linear.

\section{Description of simulation\label{sec:simulation_results}}

We are now turning to the discussion of the simulation results and its analysis, after the introduction of the external field and the simulation setup in section \ref{sec:setup_of_investigation}.

\subsection{Motion of the electron probability density\label{sec:electron_motion}}

\begin{figure}
\includegraphics[width=0.48 \textwidth]{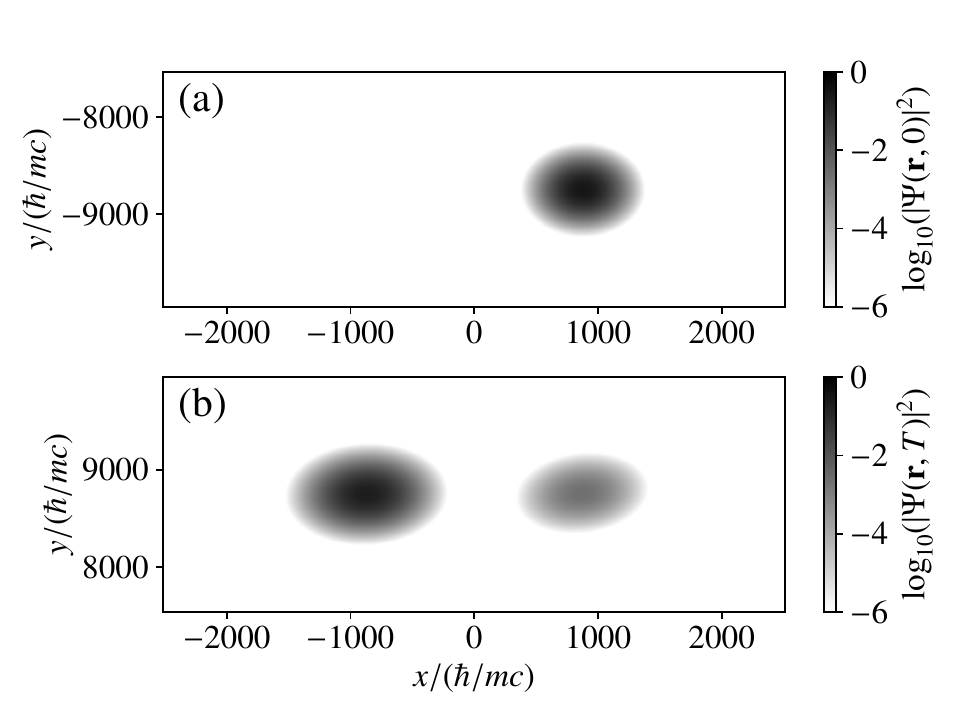}
\caption{\label{Position_of_electron} Probability density of the electron at initial and final time. The upper panel displays the initial electron probability density $|\Psi(\vec r,0)|^2$, at time $t=0$ according to Eq. \eqref{eq:probability_density}, which corresponds to the initial condition \eqref{Density_distribution_of_Gaussian_wave_packet}. The lower panel shows the electron probability $|\Psi(\vec r,T)|^2$ at final time $t=T$. Due to the co-moving simulation area, as illustrated in Fig. \ref{KDE_simulation_process}, the electron remains centered along the vertical and only moves from the right to the left. The light gray peak on the right of the lower panel is the diffracted portion of the electron wave function.}
\end{figure}

\begin{figure}
\includegraphics[width=0.48 \textwidth]{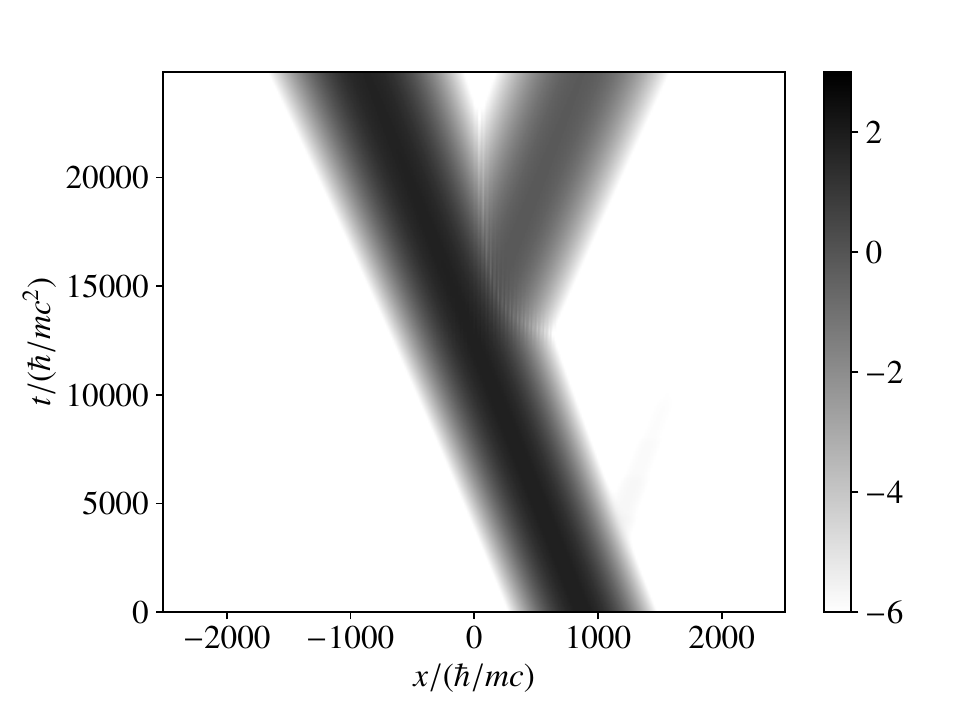}
\caption{\label{The motion_of_electron_in_X_direction}
The $y$-integrated probability density $\Phi(x,t)$ as in Eq. \eqref{eq:y_integrated_probability_density} over time. One can see that the electron moves from the right to the left, where the diffracted beam forms at the center of the figure and moves to the right, demonstrating quantum dynamics as in the Kapitza-Dirac effect. The lower leg of the $y$-shaped figure corresponds to the peak of the initial state in Fig. \ref{Position_of_electron}(a), whereas the two upper legs correspond to the two final peaks in Fig. \ref{Position_of_electron}(b).}
\end{figure}

We display the electron probability density
\begin{equation}
 |\Psi(\vec r,t)|^2 = \Psi(\vec r,t)^\dagger \Psi(\vec r,t)\label{eq:probability_density}
\end{equation}
at initial time $t=0$ in Fig. \ref{Position_of_electron}(a) and after propagation for the simulation time $T$ in Fig.  \ref{Position_of_electron}(b), where all parameters are used as described in section \ref{sec:simulation_parameters}. For illustration of the process, which takes place between the situation in Fig. \ref{Position_of_electron}(a) and Fig. \ref{Position_of_electron}(b), we compute the $y$-averaged density
\begin{equation}
 \Phi(x,t)=\int_{y_\textrm{min}(t)}^{y_\textrm{max}(t)}|\Psi(x,y,t)|^2 dy\label{eq:y_integrated_probability_density}
\end{equation}
and display it in Fig. \ref{The motion_of_electron_in_X_direction}. We observe in Fig. \ref{The motion_of_electron_in_X_direction} that the electron is moving from the right to the left, corresponding the the initially set and the expected electron positions along the $x$-axis at $x=140 \lambda_C$ and $x=-140 \lambda_C$, respectively. However, due to the interaction of the electron with the laser beam at time $T/2$, a diffracted part appears in the central region of Fig. \ref{The motion_of_electron_in_X_direction}, which moves from the center to the right, displaying the Kapitza-Dirac effect. The dynamics of the electron in Fig. \ref{The motion_of_electron_in_X_direction} explains the motion of the initial location of the electron on the right in Fig. \ref{Position_of_electron}(a) to the left in Fig. \ref{Position_of_electron}(b). Accordingly, the gray peak at  the right of Fig. \ref{Position_of_electron}(b) corresponds to the diffracted electron beam. Note, that the electron is not showing any significant motion along the $y$-axis in Fig. \ref{Position_of_electron}, as we are moving the simulation box with the electron along the $y$-direction, corresponding to the sketch in Fig. \ref{KDE_simulation_process}. An animation of the position space dynamics of the electron density $|\Psi(x,y,t)|^2$ as in Fig. \ref{Position_of_electron} is provided in the supplemental material.

\subsection{Investigation of spin resolved quantum dynamics\label{sec:spin_dynamics}}

\begin{figure} 
\includegraphics[width=0.48 \textwidth]{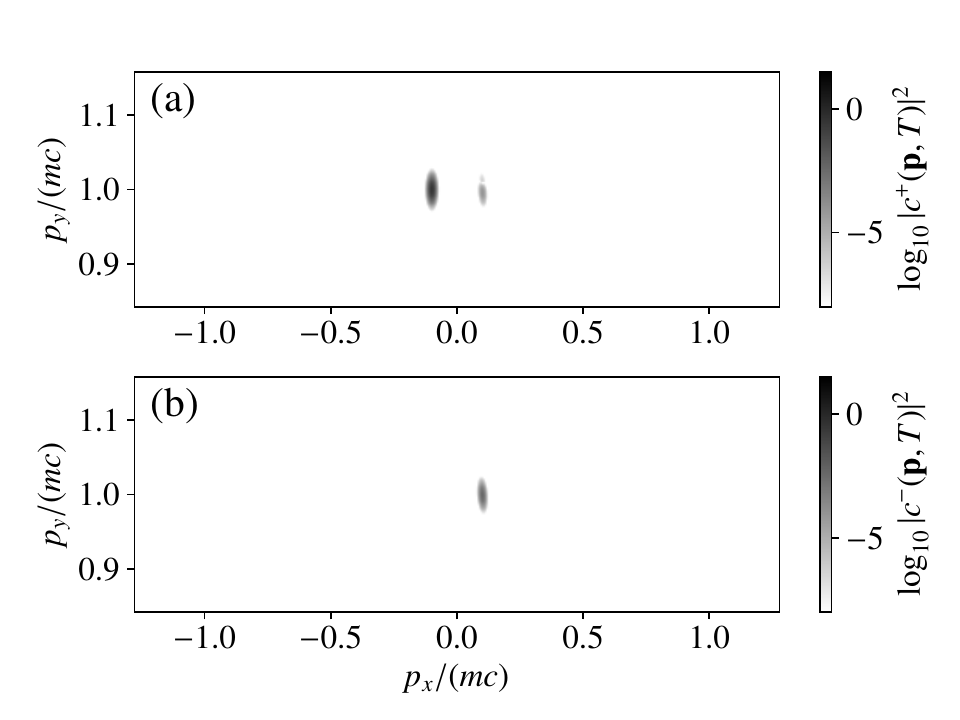}
\caption{\label{fig:spin_projection} Electron spin projection $|c^{s}(\vec p,T)|^2$ along the $x$-polarization direction according to Eq. \eqref{Equation_of_spin_projection} at the end of the simulation. The projection direction $s_f$ is positive in the upper panel and negative in the lower panel. Since the spin is determined in momentum space, this figure corresponds to the spin-resolved momentum space probability of the final electron density in Fig. \ref{Position_of_electron}(b). We observe a diffraction of the initial electron beam on the left peak of the upper panel, into a diffracted portion of the wave function, which appears two photon momenta $2 k_L \vec e_x$ to the right. The spin $-x$ component of the diffracted beam (lower panel) appears larger than the spin $+x$ component (upper panel, right peak).}
\end{figure}

Having demonstrated the quantum dynamics as predicted by the Kapitza-Dirac effect, we want to further present spin effects as discussed in \cite{ahrens_bauke_2013_relativistic_KDE,ahrens_2020_two_photon_bragg_scattering,ahrens_guan_2022_beam_focus_longitudinal}, which we display in terms of the spin-projections
\begin{equation} \label{Equation_of_spin_projection}
c^{s}(\vec{p},t) =  \braket{u^{s}(\vec p)|\varphi(\vec p,t)} \,.
\end{equation}
The absolute value squared of the transition \eqref{Equation_of_spin_projection} is displayed in Fig. \ref{fig:spin_projection} at time $T$, the end of the simulation period. Thus, Fig. \ref{fig:spin_projection} corresponds to the momentum space situation of the position space density in Fig. \ref{Position_of_electron}(b).

The prominent peak on the left in Fig. \ref{fig:spin_projection}(a) corresponds to the initial condition \eqref{Density_distribution_of_Gaussian_wave_packet} with momentum coordinate \eqref{eq:electron_momentum} and remains merely unchanged during the course of the simulation. It corresponds to the electron's motion from the right to the left in Fig. \ref{The motion_of_electron_in_X_direction}. In contrast, the right peak Fig. \ref{fig:spin_projection}(a) and the peak in Fig. \ref{fig:spin_projection}(b), arise due to the interaction of the electron with the laser, and correspond to the right moving Bragg peak in Fig. \ref{The motion_of_electron_in_X_direction}. The initial condition and the appearance of the Bragg peak over time can be viewed in detail in the animations of Fig. \ref{fig:spin_projection} in the supplemental material. The figure allows for the association of spin-polarization with the moving and diffracted portions of the electron wave function. While the left moving electron beam is purely polarized along the positive $x$ direction (as implied by the initial condition), the diffracted beam depicts a contribution with $s=+$ as well as $s=-$. Note, that the peak of the negative spin-$x$ polarization appears to be more pronounced than the peak of the positive spin-$x$ polarization.

\begin{figure}
\includegraphics[width=0.48 \textwidth]{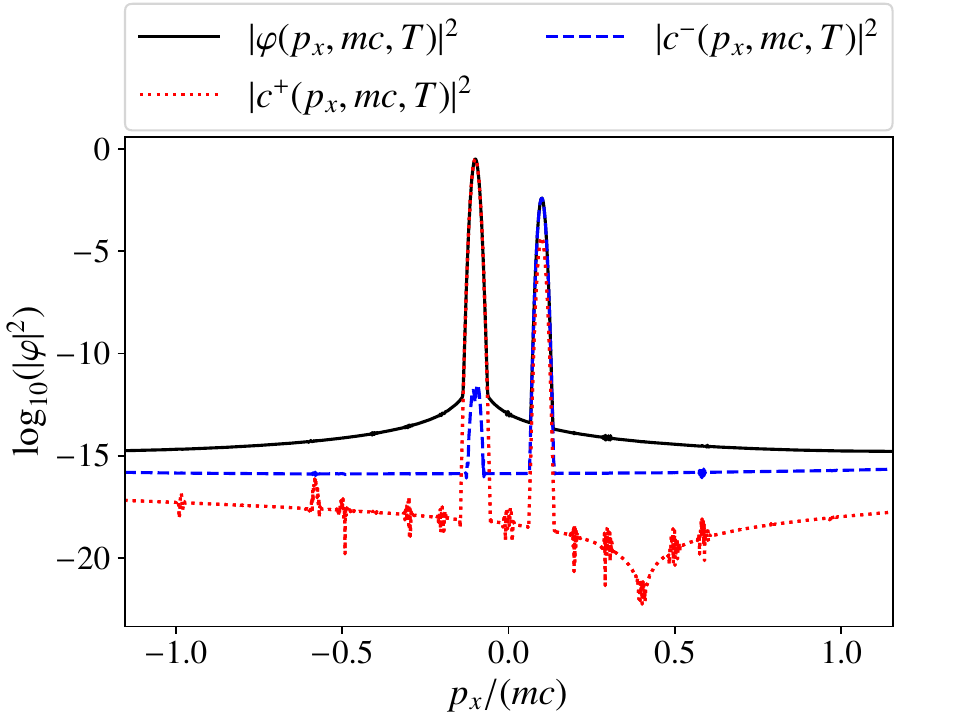}
\caption{\label{Plot_the_spin_projection}Spin resolved momentum space density of the electron along the $p_x$-axis at final simulation time $t=T$ and at $y$-momentum $p_y=mc$. Displayed are the probability density $|\varphi|^2$, according to Eq. \eqref{eq:px_momentum_space_probability_density} and the spin projections $|c^s|^2$ according to Eq. \eqref{Equation_of_spin_projection}. One can see that the momentum density $|\varphi|^2$ of the initial electron beam (left peak at $p_x=-\hbar k_L$) is coinciding with the spin $+x$ component $|c^+|^2$. However, the diffracted beam (right peak at $p_x=\hbar k_L$), is coinciding with the spin $-x$ component $|c^-|^2$. The spin $+x$ component in the diffracted beam is smaller by about two orders of magnitude and demonstrates a spin-flip in the Kapitza-Dirac effect.}
\end{figure}

For quantifying the spin amplitude, we plot the wave function's probability density in momentum space
\begin{equation}
 |\varphi(\vec p,t)|^2 = \varphi(\vec p,t)^\dagger \varphi(\vec p,t)\label{eq:px_momentum_space_probability_density}
\end{equation}
together with the spin projections \eqref{Equation_of_spin_projection} at final time $T$ at the $y$-axis position $p_y=mc$ in Fig. \ref{Plot_the_spin_projection}. We observe the initial beam on the left and the diffracted beam on the right, corresponding to the identification which we have already done in Fig. \ref{fig:spin_projection}. One can see, that the projection of the spin $+x$-polarization $|c^+|^2$ is coinciding with the probability density $|\varphi|^2$ for the initial beam. On contrary, it is the projection of the spin $-x$-polarization $|c^-|^2$, which matches the probability density  $|\varphi|^2$ of the diffracted beam. In numbers, the diffracted beam's spin $+x$ polarization amplitude is $|c^+(0.1mc,mc,T)|^2=4.6\times 10^{-5}$, whereas the spin $-x$-polarization amplitude $|c^-(0.1mc,mc,T)|^2=3.9\times 10^{-3}$ is larger by about two orders of magnitude. We thus conclude clear spin-flip dynamics along the $x$-spin polarization axis from our simulation, which agrees with the predictions in references \cite{ahrens_bauke_2013_relativistic_KDE} and \cite{ahrens_2020_two_photon_bragg_scattering}. A time-evolution of the wave function's in-field dynamics of $|\varphi(p_x,mc,t)|^2$ and $|c^s(p_x,mc,t)|^2$ in a similar fashion as in Fig. \ref{Plot_the_spin_projection} is provided in the supplemental material of this article.

\section{Analysis of physical properties from simulation\label{sec:further_analysis}}

\subsection{Validity of plane wave approximation\label{sec:plane_wave_validity}}

\begin{figure}
\includegraphics[width=0.48 \textwidth]{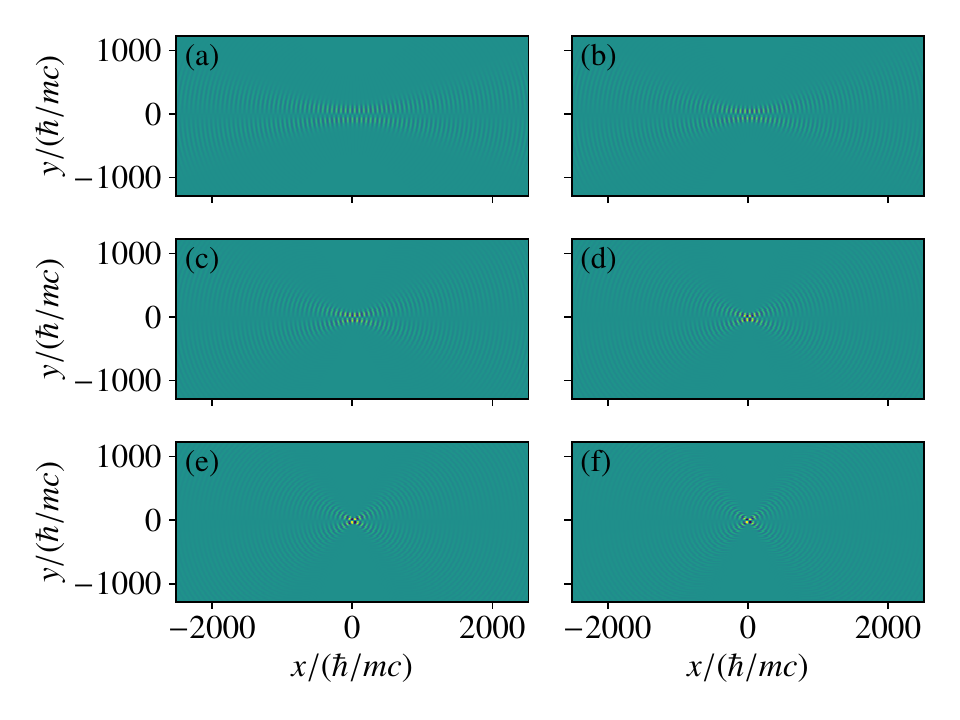}
\caption{\label{fig:A_x_variation}Longitudinal component of a Gaussian beam with increasing beam divergence $\epsilon$. The figure is arranged as Fig. \ref{fig:Simulation_of_Gaussian_beam}(a), of $A_x$ in Eq. \eqref{Equation_of_Gaussian_beam_longitudinal}. The value of $\epsilon$ in each panel is increasing according to the parameterization in Eq. \eqref{eq:increasing_epsilon_values} from panels (a) to (f), respectively. One can see that the beam is getting more focused, with increasing $\epsilon$.}
\end{figure}

\begin{figure}
\includegraphics[width=0.48 \textwidth]{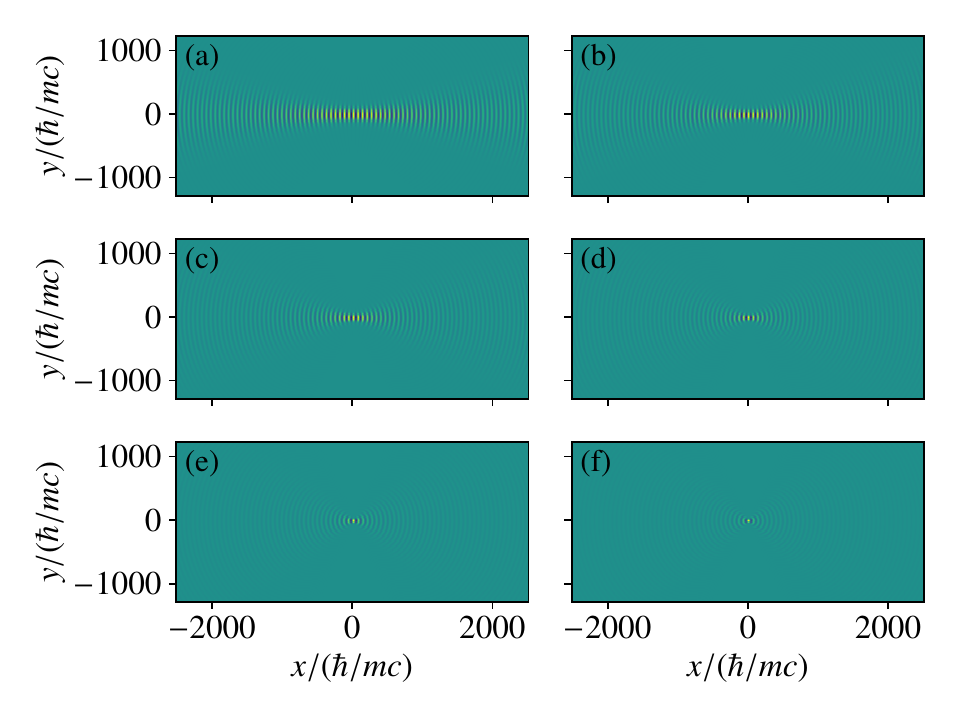}
\caption{\label{fig:A_y_variation}Transverse component of a Gaussian beam with increasing beam divergence $\epsilon$. Similarly as in Fig. \ref{fig:A_x_variation}, this figure corresponds to Fig. \ref{fig:Simulation_of_Gaussian_beam}(b), of $A_y$ in Eq. \eqref{Equation_of_Gaussian_beam_longitudinal}, with increasing $\epsilon$ values according to the parameterization in Eq. \eqref{eq:increasing_epsilon_values} from panels (a) to (f). Without the anti-symmetric zero crossing at $y=0$ of the longitudinal component, one can see the beam focusing more clearly.}
\end{figure}

The Kapitza-Dirac effect is usually described on the basis of a plane wave approximation. Our two-dimensional simulation allows us to explore, how the quantum dynamics is influenced by a strong beam focus. If one varies the beam divergence $\epsilon$ of the Gaussian beam, the stripe pattern in Fig. \ref{fig:Simulation_of_Gaussian_beam} of a nearly plane wave field turns gradually into a small focal spot. We illustrate this in Figs. \ref{fig:A_x_variation} and \ref{fig:A_y_variation} for the longitudinal ($A_x$) and transverse ($A_y$) beam polarization components, which are displayed as in Fig. \ref{fig:Simulation_of_Gaussian_beam}, but with the series of values
\begin{equation}
\epsilon = 10^{-1+q/8}\,\qquad q\in\{0,1,2,3,4,5\}\,,\label{eq:increasing_epsilon_values}
\end{equation}
for the beam divergence. All other simulation parameters are as described in section \ref{sec:simulation_results}.

\begin{figure}
\includegraphics[width=0.48 \textwidth]{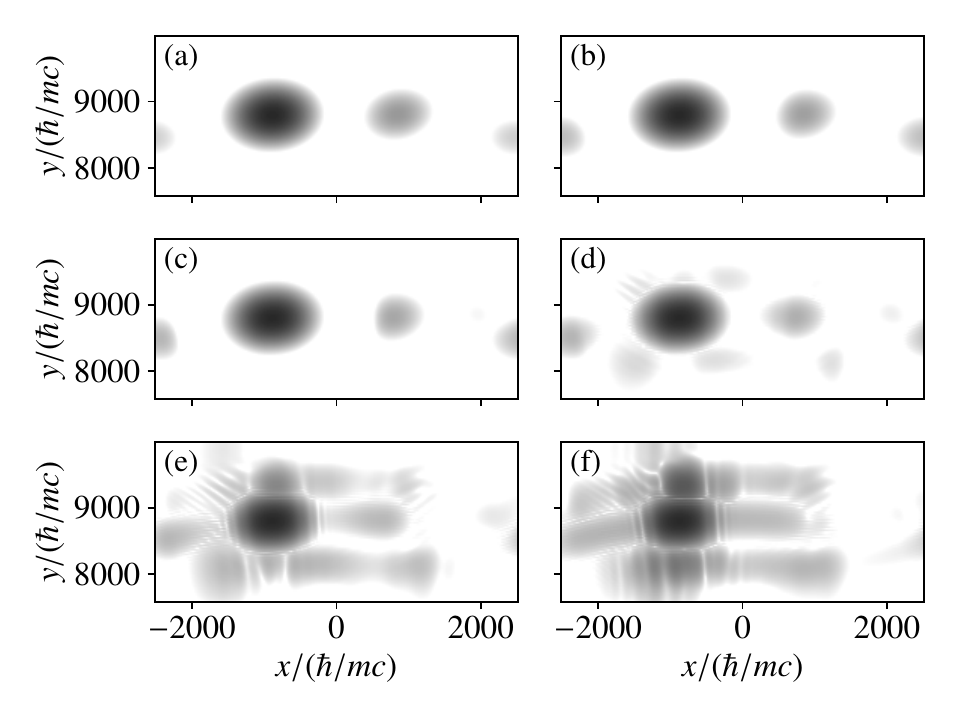}
\caption{\label{fig:position_variation}Position space diffraction probability $|\Psi(\vec r,T)|^2$ for increasing $\epsilon$. From panels (a) to (f), the beam divergence $\epsilon$ is increasing its according to Eq. \eqref{eq:increasing_epsilon_values}, in line the vector potential in Figs. \ref{fig:A_x_variation} and \ref{fig:A_y_variation}. One can see, that the plane-wave-like probability density from Fig. \ref{Position_of_electron}(b) gathers more artifacts, as the beam focus increases.}
\end{figure}

\begin{figure}
\includegraphics[width=0.48 \textwidth]{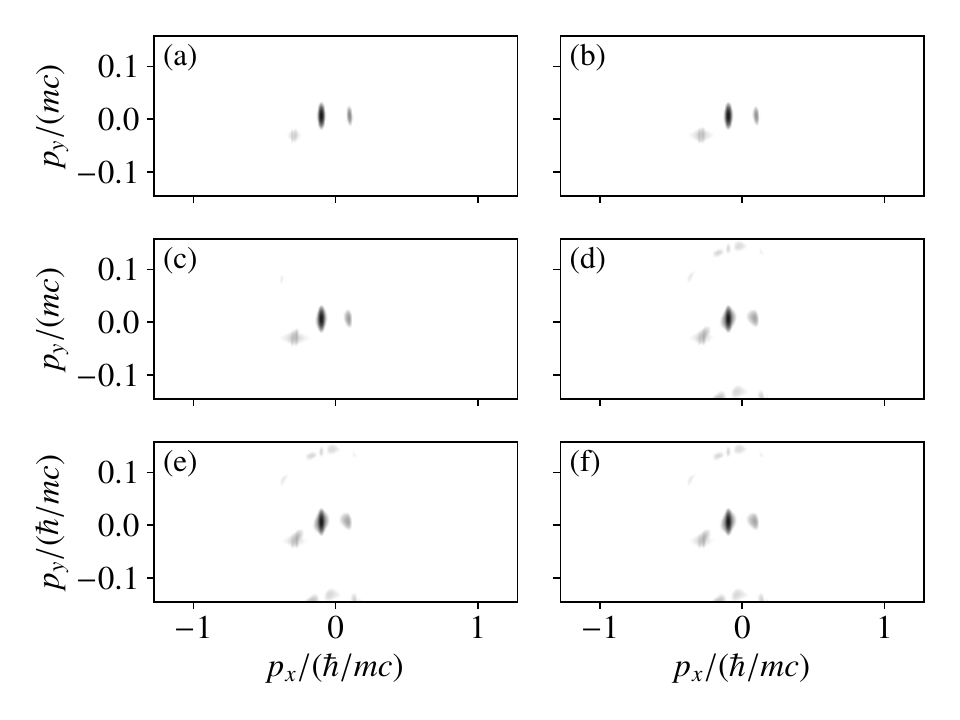}
\caption{\label{fig:momentum_variation}Momentum space diffraction probability $|\varphi(\vec r,T)|^2$ for increasing $\epsilon$. As in Figs. \ref{fig:A_x_variation} to \ref{fig:position_variation}, the beam divergence $\epsilon$ is increasing its according to Eq. \eqref{eq:increasing_epsilon_values}, from panels (a) to (f). As for the diffraction probability in position space, one can see that the plane-wave-like diffraction peaks in momentum space in Fig. \ref{fig:spin_projection} are modifying with increasing beam foci.}
\end{figure}

The effect of the beam focus on the quantum dynamics is visualized in the two-dimensional panels in Fig. \ref{fig:position_variation} in position space ($|\Psi(\vec r,T)|^2$) and Fig.  \ref{fig:momentum_variation} in momentum space ($|\varphi(\vec p,T)|^2$). In Fig. \ref{fig:position_variation} one can observe that the two peaks of the undiffracted and diffracted quantum states as they appear in Fig. \ref{Position_of_electron}(b) are surrounded by more and more artifacts, when $\epsilon$ increases. Similarly, in momentum space of Fig. \ref{fig:momentum_variation}, the two diffraction peaks which one can see in Fig. \ref{fig:spin_projection} are getting more distorted with increasing $\epsilon$, also with artifacts turning in.

\begin{figure}
\includegraphics[width=0.48 \textwidth]{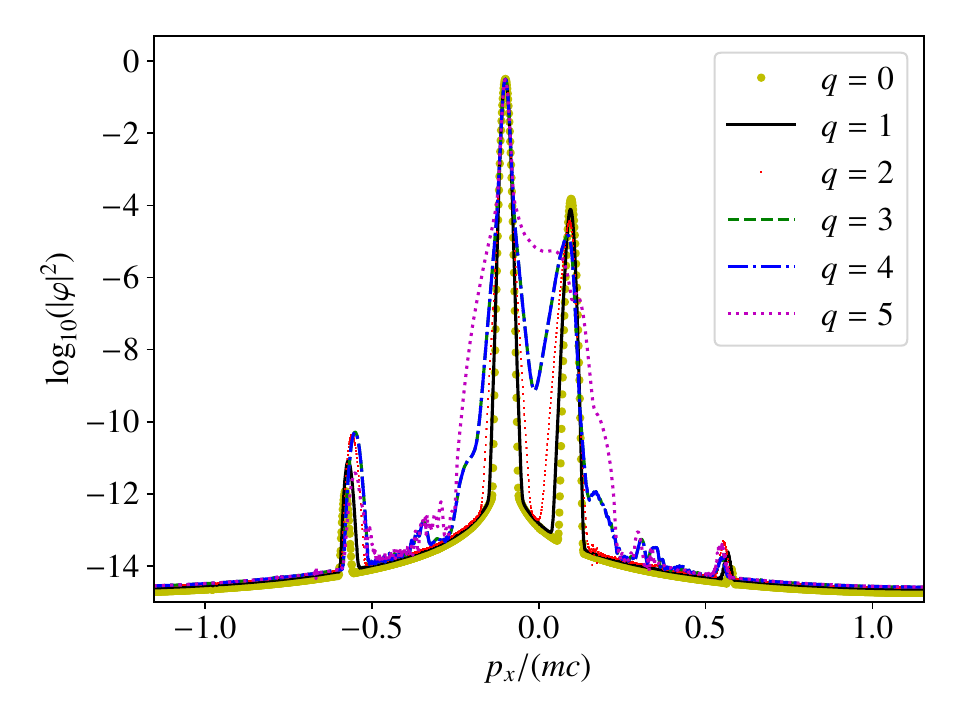}
\caption{\label{fig:peak_large_epsilon}Line plot of the momentum space diffraction probability $|\varphi(p_x,mc,T)|^2$ for increasing values of $\epsilon$. The beam divergence $\epsilon$ is varied according to Eq. \eqref{eq:increasing_epsilon_values} in a display analogous to Fig. \ref{Plot_the_spin_projection}. One can observe that the two peaks (incoming and diffracted beams) of the Kapitza-Dirac effect are broadening and merging, as $\epsilon$ increases, which illustrates the breakdown of the plane wave approximation.}
\end{figure}

For a more quantitative view on the breakdown of the plane-wave-like quantum dynamics, we display the momentum space diffraction probability $|\varphi(p_x,mc,T)|^2$ for the different values of $\epsilon$ in a line plot in Fig. \ref{fig:peak_large_epsilon}. As $\epsilon$ increases, the peaks of the incoming and diffracted peaks are broadening, until they effectively merge for $\epsilon=10^{-3/8}$. This illustrates the dismantling of the Kapitza-Dirac effect for strong beam foci and shows the limits of the plane wave approximation, which is often used for the description of the Kapitza-Dirac effect.

\subsection{Laser frequency and beam focus scaling\label{sec:scaling_effects}}

We are interested in scaling relations of the spin dynamics in the Kapitza-Dirac effect, when changing the laser frequency or the laser beam focus, within computationally accessible parameters. For that we vary the beam divergence in a similar manner as in the previous section, but with the values $\epsilon \in \{0.01,0.02,0.05,0.1\}$, well before the breakdown of the plane wave approximation. We combine the variation of $\epsilon$ with simultaneous variations of the laser photon momentum for the values $\hbar k_L/(mc) \in \{0.05,0.07,0.1\}$, such that the Bragg condition and the condition for spin-dynamics are preserved. This implies a readjustment of the initial position \eqref{eq:electron_position} and initial momentum \eqref{eq:electron_momentum} of the electron with change of the laser wavelength $\lambda=2 \pi/k_L$. Further, a change of the laser energy will only be consistent with the change of the other mentioned parameters, if also the simulation box coordinates from section \ref{sec:simulation_parameters} in terms of $\lambda$ are modified.

\begin{figure}
\includegraphics[width=0.48 \textwidth]{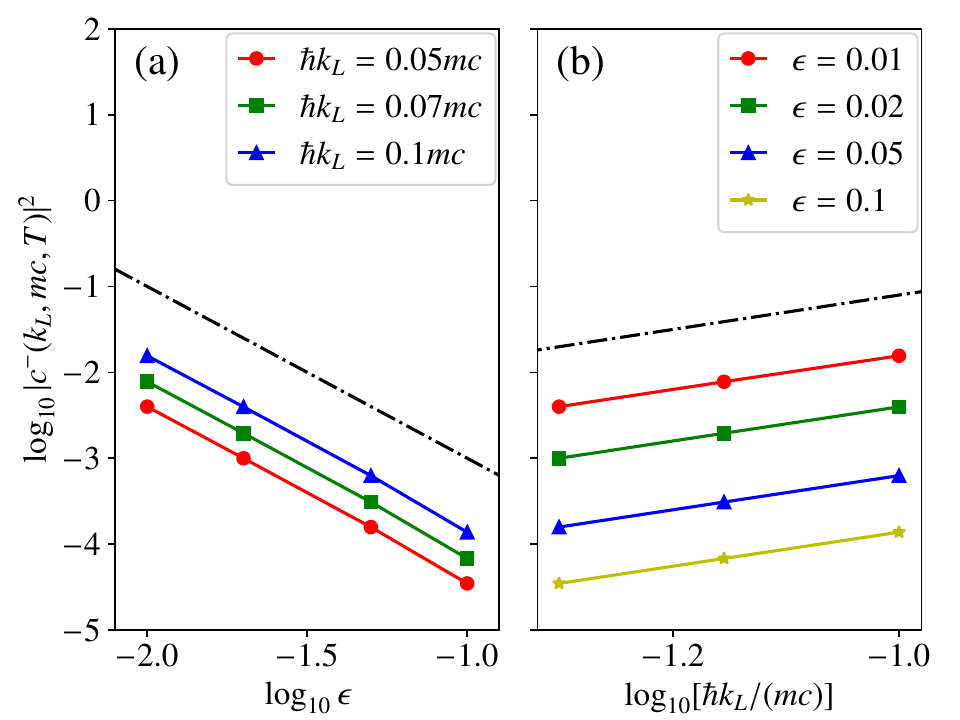}
\caption{\label{fig:spinor_projection_scaling}Spin-flip probability of spin-dynamics in the Kapitza-Dirac effect as a function of the beam divergence $\epsilon$ and the laser frequency $c k_L$. The simulation described in section \ref{sec:simulation_parameters} is varied by the parameters $\epsilon \in \{0.01,0.02,0.05,0.1\}$ and $\hbar k_L/(m c) \in \{0.05,0.07,0.1\}$, where we show the functional dependence with respect to $\epsilon$ in panel (a) and with respect to $k_L$ in panel (b). The spin-flip probability varies linearly in the double-logarithmic plot, implying a power law dependence. The black dash-dotted lines are as guide to the eye and have slope -2 in panel (a) and 2 in panel (b), respectively. We conclude a spin-flip probability scaling proportional to $(k_L/\epsilon)^2$.}
\end{figure}

The most interesting quantity be studied within the parameter variation is the spin-flip probability $|c^-(\hbar k_L,mc,T)|^2$, which corresponds to the amplitude of the diffraction peak in Fig. \ref{fig:spin_projection}(b). We display $|c^-(\hbar k_L,mc,T)|^2$ in Fig. \ref{fig:spinor_projection_scaling} for the mentioned values of $\epsilon$ and $k_L$ and see that the spin-flip probability decreases quadratically with $\epsilon$, in the double logarithmic plot in Fig. \ref{fig:spinor_projection_scaling}(a). This matches a quadratic functional dependence of the spin-flip probability with interaction time, as predicted in Eq. (22) of Ref. \cite{ahrens_2020_two_photon_bragg_scattering} due to a shorter beam crossing time through a tighter beam focus $w_0=1/(k_L \epsilon)$. In Fig. \ref{fig:spinor_projection_scaling}(b), we observe a quadratic growth of the spin-flip probability with the laser frequency $c k_L$. One would be tempted to identify the quadratic scaling in Eq. (22) of Ref. \cite{ahrens_2020_two_photon_bragg_scattering} with $k_L$ as matching property. However, since the beam waist $w_0$ also scales inversely with $k_L$, a reduced interaction time of the electron with the laser is compensating the this quadratic growth, resulting in a currently unexplained discrepancy between the wavelength scaling of the plane wave solution and the computations presented here.

\subsection{Influence of longitudinal laser polarization component\label{sec:longitudinal_component}}

\begin{figure}
\includegraphics[width=0.48 \textwidth]{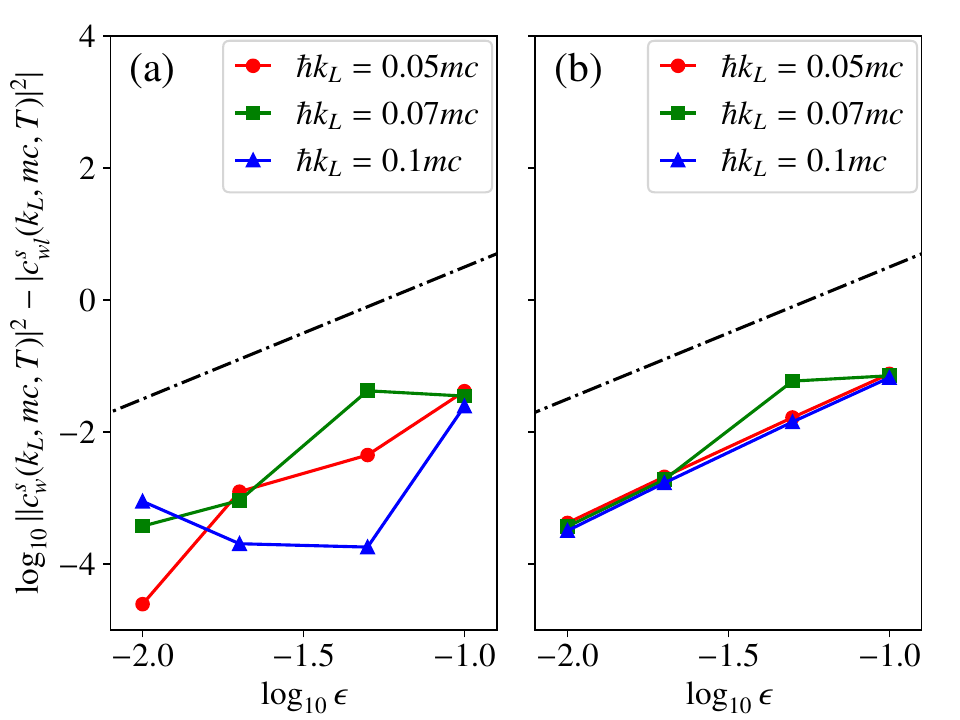}
\caption{\label{fig:spinor_projection_longitudinal}Difference of the simulated diffraction peak with longitudinal polarization component $c^s_{wl}(k_L,mc,T)$ and without $c^s_{w}(k_L,mc,T)$. Displayed are the differences $|c^s_{w}|^2-|c^s_{wl}|^2$ for the values of $\epsilon$ and $\hbar k_L$ as in Fig. \ref{fig:spinor_projection_scaling}. Panel (a) contains the spin-projection for the $+x$ polarization ($s=+$) and panel (b) the spin-projection for the $-x$ polarization ($s=-$). The black dashed dotted line is inserted for reference and has slope $2$, ie. scales as $\epsilon^2$. One can see a tendency that the difference between simulations with and without longitudinal polarization component increases quadratically with $\epsilon$.}
\end{figure}

It is further interesting to investigate the influence of the longitudinal polarization component $A_x$ of the Gaussian shaped laser beam on the quantum dynamics of the diffracted beam, which is a research question which we have investigated recently \cite{ahrens_guan_2022_beam_focus_longitudinal}. The quantum dynamics with and without the longitudinal component included appears almost identical for the values of the beam divergence $\epsilon$ and the laser photon momentum $\hbar k_L$ of the previous section. In order to quantify the scaling of the longitudinal polarization component's influence with $\epsilon$ and $\hbar k_L$, we perform our simulations with $A_x$ and with $A_x$ set to zero, and equip the spin-projections in Eq. \eqref{Equation_of_spin_projection} with an index $wl=$``with longitudinal'' [$c^s_{wl}(\vec p,t)$] and $w=$``without'' [$c^s_w(\vec p,t)$], respectively. We display the difference of the absolute value squares of $c^s_w(\vec p,t)$ and $c^s_{wl}(\vec p,t)$ in Fig. \ref{fig:spinor_projection_longitudinal}. We observe a decrease of the difference between the simulation with and without longitudinal polarization component with decreasing $\epsilon$ (quadratic scaling). This means, that the influence of the longitudinal beam polarization component on the quantum dynamics decreases for less focused laser beams. This property appears reasonable, as the longitudinal polarization component is getting smaller for less focused beams and finally vanishes for the plane wave case. We thus draw a similar conclusion as the investigation in \cite{ahrens_guan_2022_beam_focus_longitudinal}, in which the spin preserving terms from beam focusing where shown to get smaller with decreasing $\epsilon$. We also conclude from Fig. \ref{fig:spinor_projection_longitudinal}, that the influence of the longitudinal component on the quantum dynamics is rather independent of the laser photon momentum / laser wave length, which one can see particularly clear for the spin $-x$ polarization in panel (b). We attribute the more dispersed functional behavior with $\hbar k_L$ for the spin $+x$ polarization in panel (a) to resonances of other Bragg peaks and potential zero-transitions of the cross-sections, which one can also observe in Ref. \cite{ahrens_guan_2022_beam_focus_longitudinal}. Despite that, one sees the trend to a wavelength independent behaviour, in match with the results in Ref. \cite{ahrens_guan_2022_beam_focus_longitudinal}, in which some of the correction terms from the longitudinal beam polarization do not scale with $k_L$ and thus remain dominant for long laser wave lengths.

\subsection{Higher order diffraction peaks\label{sec:diffraction_regime}}

\begin{figure}
\includegraphics[width=0.48 \textwidth]{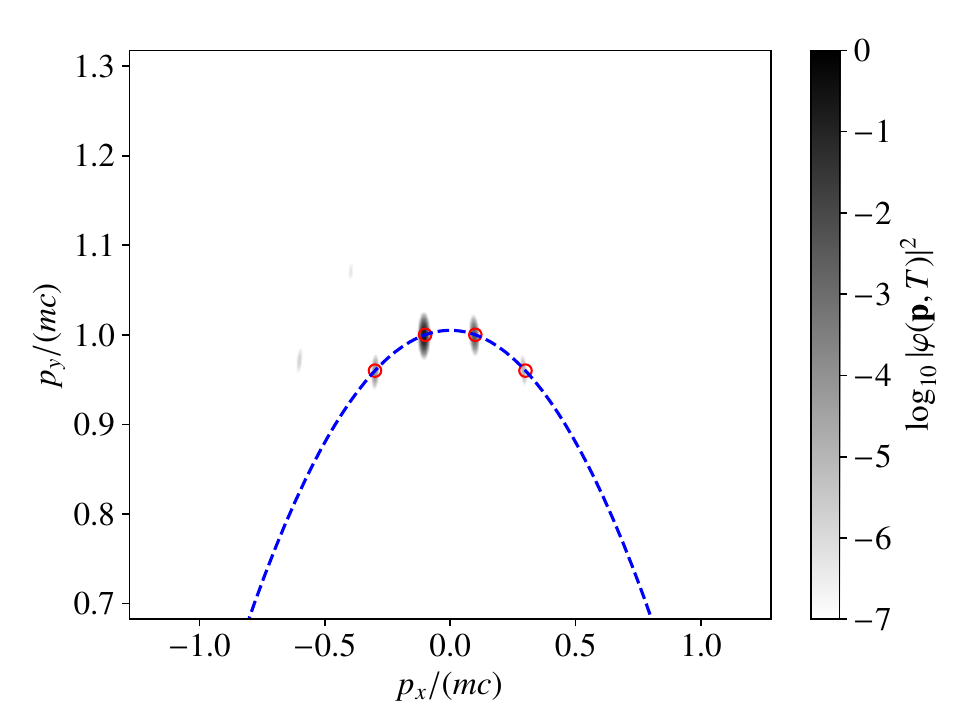}
\caption{\label{fig:momentum_density_diffraction_regime_odd}Higher order diffraction peaks in the Kapitza-Dirac effect. Displayed is the momentum space probability density for the parameters $\epsilon=0.1$ and $k_L=0.1\,mc/\hbar$, with an increased amplitude of the external laser beam vector potential $A_{0}=0.4\,mc/q$ and initial electron momentum $p_x=-\hbar k_L$ along the $x$-direction. One can see that the diffraction peaks are implied by the conservation of momentum and energy, according to Eqs. \eqref{eq:energy_momentum_conservation}. This is illustrated by the match of the diffraction peaks along the blue dashed line of Eq. \eqref{eq:equal_photon_diffraction_location}, with the red circles corresponding to $n\in \{-1,0,1,2\}$, for the case of an equal number of emitted and absorbed photons $n=n_e=n_a$.}
\end{figure}

It is possible, to couple to other diffraction orders than just the Rabi oscillations between the two momenta $\pm \hbar k_L$ in the so-called Bragg regime. Multiple diffraction orders occur in the diffraction regime, which is characterized by a tight focus of the standing light wave of the laser beam in combination with a strong amplitude of the standing light wave's ponderomotive potential \cite{batelaan_2000_KDE_first,batelaan_2007_RMP_KDE}. Therefore, in the tightly focused, short wavelength configuration with $\epsilon=0.1$ and $k_L=0.1\,mc/\hbar$, which we parameterize in the previous sections \ref{sec:scaling_effects} and \ref{sec:longitudinal_component}, we quadruple the amplitude of the laser beam's external vector potential to the value $A_{0}=0.4\,mc/q$. For reasons of numerical accuracy, we also have doubled the number of grid points along the $y$-axis to 256. We display the resulting diffraction pattern $|\varphi(\vec p,T)|^2$ of the quantum simulation with this parameter set in Fig. \ref{fig:momentum_density_diffraction_regime_odd}.

\begin{figure}
\includegraphics[width=0.48 \textwidth]{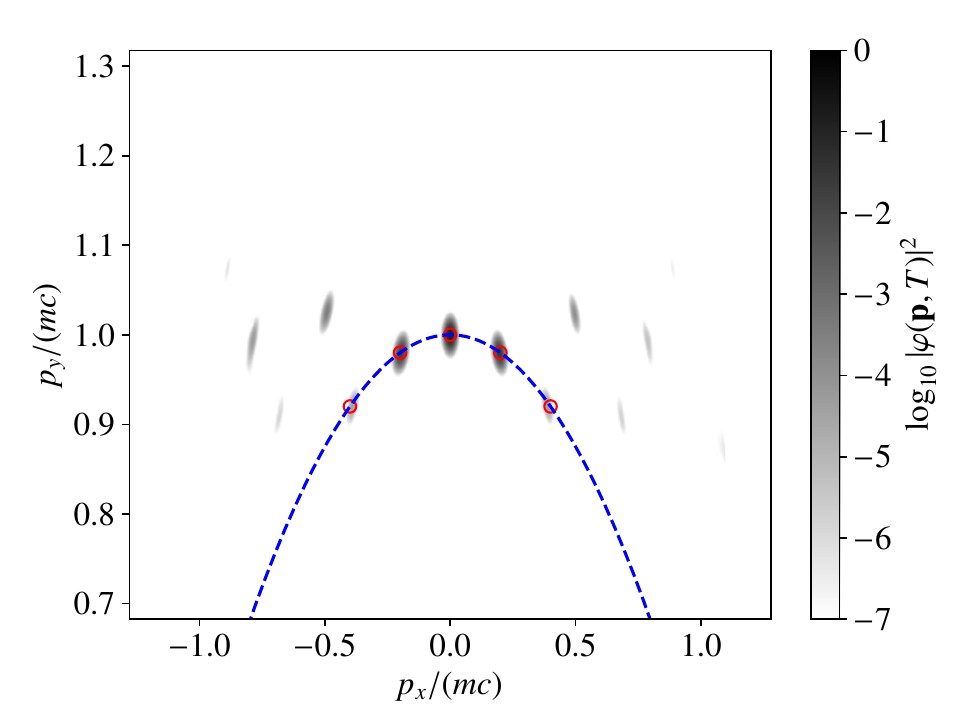}
\caption{\label{fig:momentum_density_diffraction_regime_even}Higher order diffraction peaks in the Kapitza-Dirac effect with zero initial longitudinal electron momentum ($p_x=0$). Displayed is a repetition of the simulation in Fig. \ref{fig:momentum_density_diffraction_regime_odd} with $p_x=0$. As in Fig. \ref{fig:momentum_density_diffraction_regime_odd}, the significant chain of diffraction peaks can be described by Eq. \eqref{eq:equal_photon_diffraction_location} (blue dashed line) from energy and momentum conservation \eqref{eq:energy_momentum_conservation}, with the diffraction peaks appearing for $n\in \{-2,-1,0,1,2\}$ (red circles).}
\end{figure}

We observe that the diffraction peaks are chained in a parabola-like structure, which we explain by a semi-classical argument on the basis of energy- and momentum conservation \cite{ahrens_2012_phdthesis_KDE,ahrens_bauke_2012_spin-kde,ahrens_bauke_2013_relativistic_KDE} in the following. We denote the initial and final electron momenta
\begin{subequations}\label{eq:energy_momentum_conservation}
\begin{equation}
 \vec p_{\textrm{in}} =
\begin{pmatrix}
 p_x \\ m c
\end{pmatrix}\,,\qquad
 \vec p_{\textrm{out}} =
\begin{pmatrix}
 p_x + \hbar k_L (n_a + n_e) \\ m c + \Delta p_y
\end{pmatrix}
\end{equation}
with the transverse momentum change $\Delta p_y$ and the number of absorbed/emitted photons from the left/right propagating beam $n_a$, $n_e$. Energy conservation for the electron then reads as
\begin{equation}
 E(\vec p_{\textrm{out}}) = E(\vec p_{\textrm{in}}) + \hbar c k_L (n_a - n_e)\,,\label{eq:energy_conservation}
\end{equation}
\end{subequations}
with the relativistic energy momentum relation \eqref{eq:relativistic_energy_momentum_relation}.
We expand Eq. \eqref{eq:energy_conservation} for the case of an equal number of absorbed and emitted photons $n=n_e=n_a$ and solve for $\Delta p_y$, which results in
\begin{equation}
 \Delta p_y(n)_\pm = - mc \pm \sqrt{m^2 c^4 - 4 n \hbar k_L p_x - 4n^2 \hbar^2 k_L^2}\,.
\end{equation}
We further expand the physically relevant, positive solution branch of the square root in a power series around the value $m c$ up to second order in $n$ into
\begin{equation}
 \Delta p_y(n)_+ = - 2 n \hbar k_L p_x - 2 n^2 \frac{\hbar^2 k_L^2}{m c}\,.\label{eq:equal_photon_diffraction_location}
\end{equation}
The function $\Delta p_y(n)_+$ with $p_x=-\hbar k_L$ is displayed as a blue dashed line in Fig. \ref{fig:momentum_density_diffraction_regime_even}, with the values $n\in \{-1,0,1,2\}$ marked as red circles. We conclude that the parabolically arranged chain of diffraction peaks is implied by the classical conservation of energy and momentum.

For the red circled diffraction peaks in Fig. \ref{fig:momentum_density_diffraction_regime_odd} we list the spin-resolved maximum diffraction amplitude $c^s(\vec p,T)$ as introduced in Eq. \eqref{Equation_of_spin_projection} in Table \ref{tab:diffraction_amplitudes_odd}. Despite a significant polarization along the $+x$-direction for the undiffracted beam at $n=0$, we find indications for spin-flipped beams in the neighboring states at $n=1$ and $n=-1$.

\begin{table}
\caption{
Spin-resolved amplitude of higher order diffraction peaks with initial electron momentum $p_x=-\hbar k_L$. For the diffraction peaks which are located at the red circle position in Fig. \ref{fig:momentum_density_diffraction_regime_odd}, we list the local maximum value of the spin-projections $|c^s(\vec p,T)|^2$. While the initial state at $n=0$ is strongly polarized along $+x$-direction, one can see the indication for a spin flip in the neighboring states at $n=1$ and $n=-1$.
\label{tab:diffraction_amplitudes_odd}}
\begin{tabular}{r r r}
 \hline \hline
 $n$ & \qquad$|c^+(\vec p,T)|^2$ & \qquad$|c^-(\vec p,T)|^2$ \\
 \hline
- 1 &
$1.17\times10^{-5}$ &
$1.14\times10^{-4}$ \\
0 &
$3.43\times10^{-1}$ &
$2.58\times10^{-6}$ \\
1 &
$1.53\times10^{-3}$ &
$5.17\times10^{-3}$ \\
2 &
$1.75\times10^{-5}$ &
$6.51\times10^{-6}$ \\
 \hline \hline
\end{tabular}
\end{table}

We mention additionally to the presented diffraction geometry, that the assumption of a vanishing longitudinal momentum component of the incident electron beam is a property for characterizing the diffraction regime \cite{batelaan_2000_KDE_first,batelaan_2007_RMP_KDE}. We therefore present another simulation in Fig. \ref{fig:momentum_density_diffraction_regime_even} with the same parameters as in Fig. \ref{fig:momentum_density_diffraction_regime_odd}, but with the initial longitudinal electron momentum set to zero ($p_x=0$). In Fig. \ref{fig:momentum_density_diffraction_regime_even} we also find a chain of diffraction peaks, which coincide with the condition \eqref{eq:equal_photon_diffraction_location} from energy and momentum conservation \eqref{eq:energy_momentum_conservation}, for $n\in \{-2,-1,0,1,2\}$. Further diffraction diffraction peaks are visible, which all appear at integer multiples of $\hbar k_L$. We attribute these other diffraction peaks to quantum dynamics, in which the number of emitted and absorbed photons in Eqs. \eqref{eq:energy_momentum_conservation} is not equal.

The spin-projections $c^s(\vec p,T)|^2$ of the red circled diffraction peaks in Fig. \ref{fig:momentum_density_diffraction_regime_even} are listed in Table \ref{tab:diffraction_amplitudes_even}. We find, despite a significant polarization along the $+x$ direction for the undiffracted beam at $n=0$, that the spin-flip probability ($-x$ polarization) of the diffracted beams is about an order of magnitude lower than the unflipped ($+x$ polarization) probability.

\begin{table}
\caption{
Spin-resolved amplitude of higher order diffraction peaks for $p_x=0$. Similar to Table \ref{tab:diffraction_amplitudes_odd}, we list the local maximum value $|c^s(\vec p,T)|^2$ of the red circled diffraction peaks in Fig. \ref{fig:momentum_density_diffraction_regime_odd}. Except the $+x$-polarization of the initial electron state at $n=0$, the spin-flipped probability is about an order of magnitude lower than the not-flipped probability.
\label{tab:diffraction_amplitudes_even}}
\begin{tabular}{r r r}
 \hline \hline
 $n$ & \qquad$|c^+(\vec p,T)|^2$ & \qquad$|c^-(\vec p,T)|^2$ \\
 \hline
-2 &
$3.88\times10^{-5}$ &
$3.19\times10^{-6}$ \\
-1 &
$2.20\times10^{-2}$ &
$6.62\times10^{-3}$ \\
0 &
$1.69\times10^{-1}$ &
$1.98\times10^{-7}$ \\
1 &
$2.04\times10^{-2}$ &
$6.62\times10^{-3}$ \\
2 &
$3.88\times10^{-5}$ &
$3.19\times10^{-6}$ \\
 \hline \hline
\end{tabular}
\end{table}

%

%

\section{Conclusion and outlook\label{sec:conclusions}}

In this article, we have carried out a two-dimensional, relativistic simulation of the Kapitza-Dirac effect, by using an FFT split-operator method. The standing wave laser beam is modelled by two counterpropagating Gaussian beams and thus goes beyond the plane wave ansatz of previous investigations. Likewise, the electron wave function is implemented as a finite-size Gaussian wave packet. Within the used parameters, we are able to show a Bragg peak in the Bragg regime, which is the characteristic aspect of the Kapitza-Dirac effect. Further, we have demonstrated a spin-flip along the $x$-polarization axis of the electron spin, implying that formerly discussed spin effects are theoretically possible in Kapitza-Dirac scattering, which we conclude without applying approximations.

In a subsequent study, we have investigated the breakdown of the Kapitza-Dirac plane wave diffraction dynamics, when the standing wave laser beams are getting more and more focused. We also looked at the scaling behavior of the spin-flip quantum dynamics of the Kapitza-Dirac effect with respect to changes of the laser frequency $c k_L$ and the beam divergence $\epsilon$, in the context of possible influence from the longitudinal laser polarization component from beam focusing. We conclude, that this longitudinal influence is approximately independent from the laser frequency, but increases with increasing beam divergence. Within the parameter range of our simulations, we conclude that the longitudinal polarization component will only play a significant role for the quantum dynamics for tightly focused beams, corresponding to a similar conclusion, which we were drawing in a perturbative analysis \cite{ahrens_guan_2022_beam_focus_longitudinal}. We also observe that some of the spin-dynamics in the Kapitza-Dirac effect remains present when changing the simulation parameters towards the diffraction regime.

Further investigations, which might be of interest in two-dimensional Kapitza-Dirac scattering in the future might focus on the role of negative solutions in relativistic quantum dynamics and their behavior in additional external fields.

\begin{acknowledgments}
We thank Heiko Bauke for providing the Q-Wave library for implementing quantum computations. The work was supported by the National Natural Science Foundation of China (Grants No. 11975155 and 11935008).
\end{acknowledgments}

%

\bibliography{bibliography}

\begin{thebibliography}{39}%
\makeatletter
\providecommand \@ifxundefined [1]{%
 \@ifx{#1\undefined}
}%
\providecommand \@ifnum [1]{%
 \ifnum #1\expandafter \@firstoftwo
 \else \expandafter \@secondoftwo
 \fi
}%
\providecommand \@ifx [1]{%
 \ifx #1\expandafter \@firstoftwo
 \else \expandafter \@secondoftwo
 \fi
}%
\providecommand \natexlab [1]{#1}%
\providecommand \enquote  [1]{``#1''}%
\providecommand \bibnamefont  [1]{#1}%
\providecommand \bibfnamefont [1]{#1}%
\providecommand \citenamefont [1]{#1}%
\providecommand \href@noop [0]{\@secondoftwo}%
\providecommand \href [0]{\begingroup \@sanitize@url \@href}%
\providecommand \@href[1]{\@@startlink{#1}\@@href}%
\providecommand \@@href[1]{\endgroup#1\@@endlink}%
\providecommand \@sanitize@url [0]{\catcode `\\12\catcode `\$12\catcode
  `\&12\catcode `\#12\catcode `\^12\catcode `\_12\catcode `\%12\relax}%
\providecommand \@@startlink[1]{}%
\providecommand \@@endlink[0]{}%
\providecommand \url  [0]{\begingroup\@sanitize@url \@url }%
\providecommand \@url [1]{\endgroup\@href {#1}{\urlprefix }}%
\providecommand \urlprefix  [0]{URL }%
\providecommand \Eprint [0]{\href }%
\providecommand \doibase [0]{https://doi.org/}%
\providecommand \selectlanguage [0]{\@gobble}%
\providecommand \bibinfo  [0]{\@secondoftwo}%
\providecommand \bibfield  [0]{\@secondoftwo}%
\providecommand \translation [1]{[#1]}%
\providecommand \BibitemOpen [0]{}%
\providecommand \bibitemStop [0]{}%
\providecommand \bibitemNoStop [0]{.\EOS\space}%
\providecommand \EOS [0]{\spacefactor3000\relax}%
\providecommand \BibitemShut  [1]{\csname bibitem#1\endcsname}%
\let\auto@bib@innerbib\@empty
\bibitem [{\citenamefont {Chen}\ \emph {et~al.}(2019)\citenamefont {Chen},
  \citenamefont {He}, \citenamefont {Shaisultanov}, \citenamefont
  {Hatsagortsyan},\ and\ \citenamefont
  {Keitel}}]{Chen_Keitel_2019_polarized_positrons}%
  \BibitemOpen
  \bibfield  {author} {\bibinfo {author} {\bibfnamefont {Y.-Y.}\ \bibnamefont
  {Chen}}, \bibinfo {author} {\bibfnamefont {P.-L.}\ \bibnamefont {He}},
  \bibinfo {author} {\bibfnamefont {R.}~\bibnamefont {Shaisultanov}}, \bibinfo
  {author} {\bibfnamefont {K.~Z.}\ \bibnamefont {Hatsagortsyan}},\ and\
  \bibinfo {author} {\bibfnamefont {C.~H.}\ \bibnamefont {Keitel}},\ }\bibfield
   {title} {\bibinfo {title} {{P}olarized {P}ositron {B}eams via {I}ntense
  {T}wo-{C}olor {L}aser {P}ulses},\ }\href
  {https://doi.org/10.1103/PhysRevLett.123.174801} {\bibfield  {journal}
  {\bibinfo  {journal} {Phys. Rev. Lett.}\ }\textbf {\bibinfo {volume} {123}},\
  \bibinfo {pages} {174801} (\bibinfo {year} {2019})}\BibitemShut {NoStop}%
\bibitem [{\citenamefont {Li}\ \emph {et~al.}(2019)\citenamefont {Li},
  \citenamefont {Shaisultanov}, \citenamefont {Hatsagortsyan}, \citenamefont
  {Wan}, \citenamefont {Keitel},\ and\ \citenamefont
  {Li}}]{Li_Keitel_Li_2019_single_shot_polarization}%
  \BibitemOpen
  \bibfield  {author} {\bibinfo {author} {\bibfnamefont {Y.-F.}\ \bibnamefont
  {Li}}, \bibinfo {author} {\bibfnamefont {R.}~\bibnamefont {Shaisultanov}},
  \bibinfo {author} {\bibfnamefont {K.~Z.}\ \bibnamefont {Hatsagortsyan}},
  \bibinfo {author} {\bibfnamefont {F.}~\bibnamefont {Wan}}, \bibinfo {author}
  {\bibfnamefont {C.~H.}\ \bibnamefont {Keitel}},\ and\ \bibinfo {author}
  {\bibfnamefont {J.-X.}\ \bibnamefont {Li}},\ }\bibfield  {title} {\bibinfo
  {title} {{U}ltrarelativistic {E}lectron-{B}eam {P}olarization in
  {S}ingle-{S}hot {I}nteraction with an {U}ltraintense {L}aser {P}ulse},\
  }\href {https://doi.org/10.1103/PhysRevLett.122.154801} {\bibfield  {journal}
  {\bibinfo  {journal} {Phys. Rev. Lett.}\ }\textbf {\bibinfo {volume} {122}},\
  \bibinfo {pages} {154801} (\bibinfo {year} {2019})}\BibitemShut {NoStop}%
\bibitem [{\citenamefont {Li}\ \emph {et~al.}(2020)\citenamefont {Li},
  \citenamefont {Chen}, \citenamefont {Wang},\ and\ \citenamefont
  {Hu}}]{Li_Chen_2020_polarized_positron_electron}%
  \BibitemOpen
  \bibfield  {author} {\bibinfo {author} {\bibfnamefont {Y.-F.}\ \bibnamefont
  {Li}}, \bibinfo {author} {\bibfnamefont {Y.-Y.}\ \bibnamefont {Chen}},
  \bibinfo {author} {\bibfnamefont {W.-M.}\ \bibnamefont {Wang}},\ and\
  \bibinfo {author} {\bibfnamefont {H.-S.}\ \bibnamefont {Hu}},\ }\bibfield
  {title} {\bibinfo {title} {{P}roduction of {H}ighly {P}olarized {P}ositron
  {B}eams via {H}elicity {T}ransfer from {P}olarized {E}lectrons in a {S}trong
  {L}aser {F}ield},\ }\href {https://doi.org/10.1103/PhysRevLett.125.044802}
  {\bibfield  {journal} {\bibinfo  {journal} {Phys. Rev. Lett.}\ }\textbf
  {\bibinfo {volume} {125}},\ \bibinfo {pages} {044802} (\bibinfo {year}
  {2020})}\BibitemShut {NoStop}%
\bibitem [{\citenamefont {Wen}\ \emph {et~al.}(2019)\citenamefont {Wen},
  \citenamefont {Tamburini},\ and\ \citenamefont
  {Keitel}}]{Wen_Tamburini_Keitel_2019_polarized_kilo_ampere_beams}%
  \BibitemOpen
  \bibfield  {author} {\bibinfo {author} {\bibfnamefont {M.}~\bibnamefont
  {Wen}}, \bibinfo {author} {\bibfnamefont {M.}~\bibnamefont {Tamburini}},\
  and\ \bibinfo {author} {\bibfnamefont {C.~H.}\ \bibnamefont {Keitel}},\
  }\bibfield  {title} {\bibinfo {title} {{P}olarized
  {L}aser-{W}ake{F}ield-{A}ccelerated {K}iloampere {E}lectron {B}eams},\ }\href
  {https://doi.org/10.1103/PhysRevLett.122.214801} {\bibfield  {journal}
  {\bibinfo  {journal} {Phys. Rev. Lett.}\ }\textbf {\bibinfo {volume} {122}},\
  \bibinfo {pages} {214801} (\bibinfo {year} {2019})}\BibitemShut {NoStop}%
\bibitem [{\citenamefont {Del~Sorbo}\ \emph {et~al.}(2017)\citenamefont
  {Del~Sorbo}, \citenamefont {Seipt}, \citenamefont {Blackburn}, \citenamefont
  {Thomas}, \citenamefont {Murphy}, \citenamefont {Kirk},\ and\ \citenamefont
  {Ridgers}}]{Del_Seipt_Blackburn_Kirk_2017_electron_spin_polarization}%
  \BibitemOpen
  \bibfield  {author} {\bibinfo {author} {\bibfnamefont {D.}~\bibnamefont
  {Del~Sorbo}}, \bibinfo {author} {\bibfnamefont {D.}~\bibnamefont {Seipt}},
  \bibinfo {author} {\bibfnamefont {T.~G.}\ \bibnamefont {Blackburn}}, \bibinfo
  {author} {\bibfnamefont {A.~G.~R.}\ \bibnamefont {Thomas}}, \bibinfo {author}
  {\bibfnamefont {C.~D.}\ \bibnamefont {Murphy}}, \bibinfo {author}
  {\bibfnamefont {J.~G.}\ \bibnamefont {Kirk}},\ and\ \bibinfo {author}
  {\bibfnamefont {C.~P.}\ \bibnamefont {Ridgers}},\ }\bibfield  {title}
  {\bibinfo {title} {Spin polarization of electrons by ultraintense lasers},\
  }\href {https://doi.org/10.1103/PhysRevA.96.043407} {\bibfield  {journal}
  {\bibinfo  {journal} {Phys. Rev. A}\ }\textbf {\bibinfo {volume} {96}},\
  \bibinfo {pages} {043407} (\bibinfo {year} {2017})}\BibitemShut {NoStop}%
\bibitem [{\citenamefont
  {Karlovets}(2011)}]{Karlovets_2011_radiative_polarization}%
  \BibitemOpen
  \bibfield  {author} {\bibinfo {author} {\bibfnamefont {D.~V.}\ \bibnamefont
  {Karlovets}},\ }\bibfield  {title} {\bibinfo {title} {Radiative polarization
  of electrons in a strong laser wave},\ }\href
  {https://doi.org/10.1103/PhysRevA.84.062116} {\bibfield  {journal} {\bibinfo
  {journal} {Phys. Rev. A}\ }\textbf {\bibinfo {volume} {84}},\ \bibinfo
  {pages} {062116} (\bibinfo {year} {2011})}\BibitemShut {NoStop}%
\bibitem [{\citenamefont {van Kruining}\ \emph {et~al.}(2019)\citenamefont {van
  Kruining}, \citenamefont {Mackenroth},\ and\ \citenamefont
  {G\"otte}}]{van_Kruining_Mackenroth_2019_magnetic_field_polarization}%
  \BibitemOpen
  \bibfield  {author} {\bibinfo {author} {\bibfnamefont {K.}~\bibnamefont {van
  Kruining}}, \bibinfo {author} {\bibfnamefont {F.}~\bibnamefont
  {Mackenroth}},\ and\ \bibinfo {author} {\bibfnamefont {J.~B.}\ \bibnamefont
  {G\"otte}},\ }\bibfield  {title} {\bibinfo {title} {Radiative spin
  polarization of electrons in an ultrastrong magnetic field},\ }\href
  {https://doi.org/10.1103/PhysRevD.100.056014} {\bibfield  {journal} {\bibinfo
   {journal} {Phys. Rev. D}\ }\textbf {\bibinfo {volume} {100}},\ \bibinfo
  {pages} {056014} (\bibinfo {year} {2019})}\BibitemShut {NoStop}%
\bibitem [{\citenamefont {Li}\ \emph {et~al.}(2023)\citenamefont {Li},
  \citenamefont {Chen}, \citenamefont {Hatsagortsyan}, \citenamefont
  {Di~Piazza}, \citenamefont {Tamburini},\ and\ \citenamefont
  {Keitel}}]{Keitel_2023_one_loop_polarization}%
  \BibitemOpen
  \bibfield  {author} {\bibinfo {author} {\bibfnamefont {Y.-F.}\ \bibnamefont
  {Li}}, \bibinfo {author} {\bibfnamefont {Y.-Y.}\ \bibnamefont {Chen}},
  \bibinfo {author} {\bibfnamefont {K.~Z.}\ \bibnamefont {Hatsagortsyan}},
  \bibinfo {author} {\bibfnamefont {A.}~\bibnamefont {Di~Piazza}}, \bibinfo
  {author} {\bibfnamefont {M.}~\bibnamefont {Tamburini}},\ and\ \bibinfo
  {author} {\bibfnamefont {C.~H.}\ \bibnamefont {Keitel}},\ }\bibfield  {title}
  {\bibinfo {title} {{S}trong signature of one-loop self-energy in polarization
  resolved nonlinear {C}ompton scattering},\ }\href
  {https://doi.org/10.1103/PhysRevD.107.116020} {\bibfield  {journal} {\bibinfo
   {journal} {Phys. Rev. D}\ }\textbf {\bibinfo {volume} {107}},\ \bibinfo
  {pages} {116020} (\bibinfo {year} {2023})}\BibitemShut {NoStop}%
\bibitem [{\citenamefont {Kapitza}\ and\ \citenamefont
  {Dirac}(1933)}]{kapitza_dirac_1933_proposal}%
  \BibitemOpen
  \bibfield  {author} {\bibinfo {author} {\bibfnamefont {P.~L.}\ \bibnamefont
  {Kapitza}}\ and\ \bibinfo {author} {\bibfnamefont {P.~A.~M.}\ \bibnamefont
  {Dirac}},\ }\bibfield  {title} {\bibinfo {title} {{T}he reflection of
  electrons from standing light waves},\ }\href
  {https://doi.org/10.1017/S0305004100011105} {\bibfield  {journal} {\bibinfo
  {journal} {Math. Proc. Cambridge Philos. Soc.}\ }\textbf {\bibinfo {volume}
  {29}},\ \bibinfo {pages} {297} (\bibinfo {year} {1933})}\BibitemShut
  {NoStop}%
\bibitem [{\citenamefont {Freimund}\ \emph {et~al.}(2001)\citenamefont
  {Freimund}, \citenamefont {Aflatooni},\ and\ \citenamefont
  {Batelaan}}]{Freimund_Batelaan_2001_KDE_first}%
  \BibitemOpen
  \bibfield  {author} {\bibinfo {author} {\bibfnamefont {D.~L.}\ \bibnamefont
  {Freimund}}, \bibinfo {author} {\bibfnamefont {K.}~\bibnamefont
  {Aflatooni}},\ and\ \bibinfo {author} {\bibfnamefont {H.}~\bibnamefont
  {Batelaan}},\ }\bibfield  {title} {\bibinfo {title} {{O}bservation of the
  {K}apitza-{D}irac effect},\ }\href {https://doi.org/10.1038/35093065}
  {\bibfield  {journal} {\bibinfo  {journal} {Nature (London)}\ }\textbf
  {\bibinfo {volume} {413}},\ \bibinfo {pages} {142} (\bibinfo {year}
  {2001})}\BibitemShut {NoStop}%
\bibitem [{\citenamefont {Ahrens}\ \emph {et~al.}(2012)\citenamefont {Ahrens},
  \citenamefont {Bauke}, \citenamefont {Keitel},\ and\ \citenamefont
  {M\"uller}}]{ahrens_bauke_2012_spin-kde}%
  \BibitemOpen
  \bibfield  {author} {\bibinfo {author} {\bibfnamefont {S.}~\bibnamefont
  {Ahrens}}, \bibinfo {author} {\bibfnamefont {H.}~\bibnamefont {Bauke}},
  \bibinfo {author} {\bibfnamefont {C.~H.}\ \bibnamefont {Keitel}},\ and\
  \bibinfo {author} {\bibfnamefont {C.}~\bibnamefont {M\"uller}},\ }\bibfield
  {title} {\bibinfo {title} {{S}pin {D}ynamics in the {K}apitza-{D}irac
  {E}ffect},\ }\href {https://doi.org/10.1103/PhysRevLett.109.043601}
  {\bibfield  {journal} {\bibinfo  {journal} {Phys. Rev. Lett.}\ }\textbf
  {\bibinfo {volume} {109}},\ \bibinfo {pages} {043601} (\bibinfo {year}
  {2012})}\BibitemShut {NoStop}%
\bibitem [{\citenamefont {Ahrens}\ \emph {et~al.}(2013)\citenamefont {Ahrens},
  \citenamefont {Bauke}, \citenamefont {Keitel},\ and\ \citenamefont
  {M\"uller}}]{ahrens_bauke_2013_relativistic_KDE}%
  \BibitemOpen
  \bibfield  {author} {\bibinfo {author} {\bibfnamefont {S.}~\bibnamefont
  {Ahrens}}, \bibinfo {author} {\bibfnamefont {H.}~\bibnamefont {Bauke}},
  \bibinfo {author} {\bibfnamefont {C.~H.}\ \bibnamefont {Keitel}},\ and\
  \bibinfo {author} {\bibfnamefont {C.}~\bibnamefont {M\"uller}},\ }\bibfield
  {title} {\bibinfo {title} {{K}apitza-{D}irac effect in the relativistic
  regime},\ }\href {https://doi.org/10.1103/PhysRevA.88.012115} {\bibfield
  {journal} {\bibinfo  {journal} {Phys. Rev. A}\ }\textbf {\bibinfo {volume}
  {88}},\ \bibinfo {pages} {012115} (\bibinfo {year} {2013})}\BibitemShut
  {NoStop}%
\bibitem [{\citenamefont {Dellweg}\ \emph {et~al.}(2016)\citenamefont
  {Dellweg}, \citenamefont {Awwad},\ and\ \citenamefont
  {M\"uller}}]{dellweg_awwad_mueller_2016_spin-dynamics_bichromatic_laser_fields}%
  \BibitemOpen
  \bibfield  {author} {\bibinfo {author} {\bibfnamefont {M.~M.}\ \bibnamefont
  {Dellweg}}, \bibinfo {author} {\bibfnamefont {H.~M.}\ \bibnamefont {Awwad}},\
  and\ \bibinfo {author} {\bibfnamefont {C.}~\bibnamefont {M\"uller}},\
  }\bibfield  {title} {\bibinfo {title} {{S}pin dynamics in {K}apitza-{D}irac
  scattering of electrons from bichromatic laser fields},\ }\href
  {https://doi.org/10.1103/PhysRevA.94.022122} {\bibfield  {journal} {\bibinfo
  {journal} {Phys. Rev. A}\ }\textbf {\bibinfo {volume} {94}},\ \bibinfo
  {pages} {022122} (\bibinfo {year} {2016})}\BibitemShut {NoStop}%
\bibitem [{\citenamefont {Erhard}\ and\ \citenamefont
  {Bauke}(2015)}]{erhard_bauke_2015_spin}%
  \BibitemOpen
  \bibfield  {author} {\bibinfo {author} {\bibfnamefont {R.}~\bibnamefont
  {Erhard}}\ and\ \bibinfo {author} {\bibfnamefont {H.}~\bibnamefont {Bauke}},\
  }\bibfield  {title} {\bibinfo {title} {{S}pin effects in {K}apitza-{D}irac
  scattering at light with elliptical polarization},\ }\href
  {https://doi.org/10.1103/PhysRevA.92.042123} {\bibfield  {journal} {\bibinfo
  {journal} {Phys. Rev. A}\ }\textbf {\bibinfo {volume} {92}},\ \bibinfo
  {pages} {042123} (\bibinfo {year} {2015})}\BibitemShut {NoStop}%
\bibitem [{\citenamefont {Bragg}\ and\ \citenamefont
  {Bragg}(1913)}]{Bragg_1913_Bragg_scattering}%
  \BibitemOpen
  \bibfield  {author} {\bibinfo {author} {\bibfnamefont {W.~H.}\ \bibnamefont
  {Bragg}}\ and\ \bibinfo {author} {\bibfnamefont {W.~L.}\ \bibnamefont
  {Bragg}},\ }\bibfield  {title} {\bibinfo {title} {{T}he reflection of
  {X}-rays by crystals},\ }\href {https://doi.org/10.1098/rspa.1913.0040}
  {\bibfield  {journal} {\bibinfo  {journal} {Proc. R. Soc. London. Series A}\
  }\textbf {\bibinfo {volume} {88}},\ \bibinfo {pages} {428} (\bibinfo {year}
  {1913})}\BibitemShut {NoStop}%
\bibitem [{\citenamefont {Batelaan}(2000)}]{batelaan_2000_KDE_first}%
  \BibitemOpen
  \bibfield  {author} {\bibinfo {author} {\bibfnamefont {H.}~\bibnamefont
  {Batelaan}},\ }\bibfield  {title} {\bibinfo {title} {{T}he {K}apitza-{D}irac
  effect},\ }\href {https://doi.org/10.1080/00107510010001220} {\bibfield
  {journal} {\bibinfo  {journal} {Contemp. Phys.}\ }\textbf {\bibinfo {volume}
  {41}},\ \bibinfo {pages} {369} (\bibinfo {year} {2000})}\BibitemShut
  {NoStop}%
\bibitem [{\citenamefont {Batelaan}(2007)}]{batelaan_2007_RMP_KDE}%
  \BibitemOpen
  \bibfield  {author} {\bibinfo {author} {\bibfnamefont {H.}~\bibnamefont
  {Batelaan}},\ }\bibfield  {title} {\bibinfo {title} {\textit{{C}olloquium} :
  {I}lluminating the {K}apitza-{D}irac effect with electron matter optics},\
  }\href {https://doi.org/10.1103/RevModPhys.79.929} {\bibfield  {journal}
  {\bibinfo  {journal} {Rev. Mod. Phys.}\ }\textbf {\bibinfo {volume} {79}},\
  \bibinfo {pages} {929} (\bibinfo {year} {2007})}\BibitemShut {NoStop}%
\bibitem [{\citenamefont {Dellweg}\ and\ \citenamefont
  {M\"uller}(2017{\natexlab{a}})}]{dellweg_mueller_2016_interferometric_spin-polarizer}%
  \BibitemOpen
  \bibfield  {author} {\bibinfo {author} {\bibfnamefont {M.~M.}\ \bibnamefont
  {Dellweg}}\ and\ \bibinfo {author} {\bibfnamefont {C.}~\bibnamefont
  {M\"uller}},\ }\bibfield  {title} {\bibinfo {title} {{S}pin-{P}olarizing
  {I}nterferometric {B}eam {S}plitter for {F}ree {E}lectrons},\ }\href
  {https://doi.org/10.1103/PhysRevLett.118.070403} {\bibfield  {journal}
  {\bibinfo  {journal} {Phys. Rev. Lett.}\ }\textbf {\bibinfo {volume} {118}},\
  \bibinfo {pages} {070403} (\bibinfo {year} {2017}{\natexlab{a}})}\BibitemShut
  {NoStop}%
\bibitem [{\citenamefont {Ahrens}(2017)}]{ahrens_2017_spin_filter}%
  \BibitemOpen
  \bibfield  {author} {\bibinfo {author} {\bibfnamefont {S.}~\bibnamefont
  {Ahrens}},\ }\bibfield  {title} {\bibinfo {title} {{E}lectron-spin filter and
  polarizer in a standing light wave},\ }\href
  {https://doi.org/10.1103/PhysRevA.96.052132} {\bibfield  {journal} {\bibinfo
  {journal} {Phys. Rev. A}\ }\textbf {\bibinfo {volume} {96}},\ \bibinfo
  {pages} {052132} (\bibinfo {year} {2017})}\BibitemShut {NoStop}%
\bibitem [{\citenamefont {Dellweg}\ and\ \citenamefont
  {M\"uller}(2017{\natexlab{b}})}]{dellweg_mueller_extended_KDE_calculations}%
  \BibitemOpen
  \bibfield  {author} {\bibinfo {author} {\bibfnamefont {M.~M.}\ \bibnamefont
  {Dellweg}}\ and\ \bibinfo {author} {\bibfnamefont {C.}~\bibnamefont
  {M\"uller}},\ }\bibfield  {title} {\bibinfo {title} {{C}ontrolling electron
  spin dynamics in bichromatic {K}apitza-{D}irac scattering by the laser field
  polarization},\ }\href {https://doi.org/10.1103/PhysRevA.95.042124}
  {\bibfield  {journal} {\bibinfo  {journal} {Phys. Rev. A}\ }\textbf {\bibinfo
  {volume} {95}},\ \bibinfo {pages} {042124} (\bibinfo {year}
  {2017}{\natexlab{b}})}\BibitemShut {NoStop}%
\bibitem [{\citenamefont {Ahrens}\ \emph {et~al.}(2020)\citenamefont {Ahrens},
  \citenamefont {Liang}, \citenamefont {\ifmmode \check{C}\else
  \v{C}\fi{}ade\ifmmode~\check{z}\else \v{z}\fi{}},\ and\ \citenamefont
  {Shen}}]{ahrens_2020_two_photon_bragg_scattering}%
  \BibitemOpen
  \bibfield  {author} {\bibinfo {author} {\bibfnamefont {S.}~\bibnamefont
  {Ahrens}}, \bibinfo {author} {\bibfnamefont {Z.}~\bibnamefont {Liang}},
  \bibinfo {author} {\bibfnamefont {T.}~\bibnamefont {\ifmmode \check{C}\else
  \v{C}\fi{}ade\ifmmode~\check{z}\else \v{z}\fi{}}},\ and\ \bibinfo {author}
  {\bibfnamefont {B.}~\bibnamefont {Shen}},\ }\bibfield  {title} {\bibinfo
  {title} {{S}pin-dependent two-photon {B}ragg scattering in the
  {K}apitza-{D}irac effect},\ }\href
  {https://doi.org/10.1103/PhysRevA.102.033106} {\bibfield  {journal} {\bibinfo
   {journal} {Phys. Rev. A}\ }\textbf {\bibinfo {volume} {102}},\ \bibinfo
  {pages} {033106} (\bibinfo {year} {2020})}\BibitemShut {NoStop}%
\bibitem [{\citenamefont {McGregor}\ \emph {et~al.}(2015)\citenamefont
  {McGregor}, \citenamefont {Huang}, \citenamefont {Shadwick},\ and\
  \citenamefont {Batelaan}}]{McGregor_Batelaan_2015_two_color_spin}%
  \BibitemOpen
  \bibfield  {author} {\bibinfo {author} {\bibfnamefont {S.}~\bibnamefont
  {McGregor}}, \bibinfo {author} {\bibfnamefont {W.~C.-W.}\ \bibnamefont
  {Huang}}, \bibinfo {author} {\bibfnamefont {B.~A.}\ \bibnamefont
  {Shadwick}},\ and\ \bibinfo {author} {\bibfnamefont {H.}~\bibnamefont
  {Batelaan}},\ }\bibfield  {title} {\bibinfo {title} {{S}pin-dependent
  two-color {K}apitza-{D}irac effects},\ }\href
  {https://doi.org/10.1103/PhysRevA.92.023834} {\bibfield  {journal} {\bibinfo
  {journal} {Phys. Rev. A}\ }\textbf {\bibinfo {volume} {92}},\ \bibinfo
  {pages} {023834} (\bibinfo {year} {2015})}\BibitemShut {NoStop}%
\bibitem [{\citenamefont {Ebadati}\ \emph {et~al.}(2018)\citenamefont
  {Ebadati}, \citenamefont {Vafaee},\ and\ \citenamefont
  {Shokri}}]{ebadati_2018_four_photon_KDE}%
  \BibitemOpen
  \bibfield  {author} {\bibinfo {author} {\bibfnamefont {A.}~\bibnamefont
  {Ebadati}}, \bibinfo {author} {\bibfnamefont {M.}~\bibnamefont {Vafaee}},\
  and\ \bibinfo {author} {\bibfnamefont {B.}~\bibnamefont {Shokri}},\
  }\bibfield  {title} {\bibinfo {title} {{F}our-photon {K}apitza-{D}irac effect
  as an electron spin filter},\ }\href
  {https://doi.org/10.1103/PhysRevA.98.032505} {\bibfield  {journal} {\bibinfo
  {journal} {Phys. Rev. A}\ }\textbf {\bibinfo {volume} {98}},\ \bibinfo
  {pages} {032505} (\bibinfo {year} {2018})}\BibitemShut {NoStop}%
\bibitem [{\citenamefont {Ebadati}\ \emph {et~al.}(2019)\citenamefont
  {Ebadati}, \citenamefont {Vafaee},\ and\ \citenamefont
  {Shokri}}]{ebadati_2019_n_photon_KDE}%
  \BibitemOpen
  \bibfield  {author} {\bibinfo {author} {\bibfnamefont {A.}~\bibnamefont
  {Ebadati}}, \bibinfo {author} {\bibfnamefont {M.}~\bibnamefont {Vafaee}},\
  and\ \bibinfo {author} {\bibfnamefont {B.}~\bibnamefont {Shokri}},\
  }\bibfield  {title} {\bibinfo {title} {{I}nvestigation of electron spin
  dynamic in the bichromatic {K}apitza-{D}irac effect via frequency ratio and
  amplitude of laser beams},\ }\href
  {https://doi.org/10.1103/PhysRevA.100.052514} {\bibfield  {journal} {\bibinfo
   {journal} {Phys. Rev. A}\ }\textbf {\bibinfo {volume} {100}},\ \bibinfo
  {pages} {052514} (\bibinfo {year} {2019})}\BibitemShut {NoStop}%
\bibitem [{\citenamefont {Gerlach}\ and\ \citenamefont
  {Stern}(1922{\natexlab{a}})}]{Stern_Gerlach_1922_1}%
  \BibitemOpen
  \bibfield  {author} {\bibinfo {author} {\bibfnamefont {W.}~\bibnamefont
  {Gerlach}}\ and\ \bibinfo {author} {\bibfnamefont {O.}~\bibnamefont
  {Stern}},\ }\bibfield  {title} {\bibinfo {title} {Der experimentelle
  {N}achweis der {R}ichtungsquantelung im {M}agnetfeld},\ }\href
  {https://doi.org/10.1007/BF01326983} {\bibfield  {journal} {\bibinfo
  {journal} {Zeitschrift f{\"u}r Physik}\ }\textbf {\bibinfo {volume} {9}},\
  \bibinfo {pages} {349} (\bibinfo {year} {1922}{\natexlab{a}})}\BibitemShut
  {NoStop}%
\bibitem [{\citenamefont {Gerlach}\ and\ \citenamefont
  {Stern}(1922{\natexlab{b}})}]{Stern_Gerlach_1922_2}%
  \BibitemOpen
  \bibfield  {author} {\bibinfo {author} {\bibfnamefont {W.}~\bibnamefont
  {Gerlach}}\ and\ \bibinfo {author} {\bibfnamefont {O.}~\bibnamefont
  {Stern}},\ }\bibfield  {title} {\bibinfo {title} {Das magnetische {M}oment
  des {S}ilberatoms},\ }\href {https://doi.org/10.1007/BF01326984} {\bibfield
  {journal} {\bibinfo  {journal} {Zeitschrift f{\"u}r Physik}\ }\textbf
  {\bibinfo {volume} {9}},\ \bibinfo {pages} {353} (\bibinfo {year}
  {1922}{\natexlab{b}})}\BibitemShut {NoStop}%
\bibitem [{\citenamefont {Gerlach}\ and\ \citenamefont
  {Stern}(1922{\natexlab{c}})}]{Stern_Gerlach_1922_3}%
  \BibitemOpen
  \bibfield  {author} {\bibinfo {author} {\bibfnamefont {W.}~\bibnamefont
  {Gerlach}}\ and\ \bibinfo {author} {\bibfnamefont {O.}~\bibnamefont
  {Stern}},\ }\bibfield  {title} {\bibinfo {title} {Der experimentelle
  {N}achweis des magnetischen {M}oments des {S}ilberatoms},\ }\href
  {https://doi.org/10.1007/BF01329580} {\bibfield  {journal} {\bibinfo
  {journal} {Zeitschrift f{\"u}r Physik}\ }\textbf {\bibinfo {volume} {8}},\
  \bibinfo {pages} {110} (\bibinfo {year} {1922}{\natexlab{c}})}\BibitemShut
  {NoStop}%
\bibitem [{\citenamefont {Compton}(1923)}]{Compton_1923_compton_scattering}%
  \BibitemOpen
  \bibfield  {author} {\bibinfo {author} {\bibfnamefont {A.~H.}\ \bibnamefont
  {Compton}},\ }\bibfield  {title} {\bibinfo {title} {{A} {Q}uantum {T}heory of
  the {S}cattering of {X}-rays by {L}ight {E}lements},\ }\href
  {https://doi.org/10.1103/PhysRev.21.483} {\bibfield  {journal} {\bibinfo
  {journal} {Phys. Rev.}\ }\textbf {\bibinfo {volume} {21}},\ \bibinfo {pages}
  {483} (\bibinfo {year} {1923})}\BibitemShut {NoStop}%
\bibitem [{\citenamefont {Freimund}\ and\ \citenamefont
  {Batelaan}(2002)}]{Freimund_Batelaan_2002_KDE_detection_PRL}%
  \BibitemOpen
  \bibfield  {author} {\bibinfo {author} {\bibfnamefont {D.~L.}\ \bibnamefont
  {Freimund}}\ and\ \bibinfo {author} {\bibfnamefont {H.}~\bibnamefont
  {Batelaan}},\ }\bibfield  {title} {\bibinfo {title} {{B}ragg {S}cattering of
  {F}ree {E}lectrons {U}sing the {K}apitza-{D}irac {E}ffect},\ }\href
  {https://doi.org/10.1103/PhysRevLett.89.283602} {\bibfield  {journal}
  {\bibinfo  {journal} {Phys. Rev. Lett.}\ }\textbf {\bibinfo {volume} {89}},\
  \bibinfo {pages} {283602} (\bibinfo {year} {2002})}\BibitemShut {NoStop}%
\bibitem [{\citenamefont {Axelrod}\ \emph {et~al.}(2020)\citenamefont
  {Axelrod}, \citenamefont {Campbell}, \citenamefont {Schwartz}, \citenamefont
  {Turnbaugh}, \citenamefont {Glaeser},\ and\ \citenamefont
  {M\"uller}}]{Axelrod_2020_Kapitza_Dirac_cancellation_observation}%
  \BibitemOpen
  \bibfield  {author} {\bibinfo {author} {\bibfnamefont {J.~J.}\ \bibnamefont
  {Axelrod}}, \bibinfo {author} {\bibfnamefont {S.~L.}\ \bibnamefont
  {Campbell}}, \bibinfo {author} {\bibfnamefont {O.}~\bibnamefont {Schwartz}},
  \bibinfo {author} {\bibfnamefont {C.}~\bibnamefont {Turnbaugh}}, \bibinfo
  {author} {\bibfnamefont {R.~M.}\ \bibnamefont {Glaeser}},\ and\ \bibinfo
  {author} {\bibfnamefont {H.}~\bibnamefont {M\"uller}},\ }\bibfield  {title}
  {\bibinfo {title} {{O}bservation of the {R}elativistic {R}eversal of the
  {P}onderomotive {P}otential},\ }\href
  {https://doi.org/10.1103/PhysRevLett.124.174801} {\bibfield  {journal}
  {\bibinfo  {journal} {Phys. Rev. Lett.}\ }\textbf {\bibinfo {volume} {124}},\
  \bibinfo {pages} {174801} (\bibinfo {year} {2020})}\BibitemShut {NoStop}%
\bibitem [{\citenamefont {Dickson}(1970)}]{Dickson_1970_gaussian_beam}%
  \BibitemOpen
  \bibfield  {author} {\bibinfo {author} {\bibfnamefont {L.~D.}\ \bibnamefont
  {Dickson}},\ }\bibfield  {title} {\bibinfo {title} {{C}haracteristics of a
  {P}ropagating {G}aussian {B}eam},\ }\href
  {https://doi.org/10.1364/AO.9.001854} {\bibfield  {journal} {\bibinfo
  {journal} {Appl. Opt.}\ }\textbf {\bibinfo {volume} {9}},\ \bibinfo {pages}
  {1854} (\bibinfo {year} {1970})}\BibitemShut {NoStop}%
\bibitem [{\citenamefont {Ahrens}\ \emph {et~al.}(2022)\citenamefont {Ahrens},
  \citenamefont {Guan},\ and\ \citenamefont
  {Shen}}]{ahrens_guan_2022_beam_focus_longitudinal}%
  \BibitemOpen
  \bibfield  {author} {\bibinfo {author} {\bibfnamefont {S.}~\bibnamefont
  {Ahrens}}, \bibinfo {author} {\bibfnamefont {Z.}~\bibnamefont {Guan}},\ and\
  \bibinfo {author} {\bibfnamefont {B.}~\bibnamefont {Shen}},\ }\bibfield
  {title} {\bibinfo {title} {Beam focus and longitudinal polarization influence
  on spin dynamics in the {K}apitza-{D}irac effect},\ }\href
  {https://doi.org/10.1103/PhysRevA.105.053123} {\bibfield  {journal} {\bibinfo
   {journal} {Phys. Rev. A}\ }\textbf {\bibinfo {volume} {105}},\ \bibinfo
  {pages} {053123} (\bibinfo {year} {2022})}\BibitemShut {NoStop}%
\bibitem [{\citenamefont {Braun}\ \emph {et~al.}(1999)\citenamefont {Braun},
  \citenamefont {Su},\ and\ \citenamefont
  {Grobe}}]{Grobe_1999_FFT_split_operator_method}%
  \BibitemOpen
  \bibfield  {author} {\bibinfo {author} {\bibfnamefont {J.~W.}\ \bibnamefont
  {Braun}}, \bibinfo {author} {\bibfnamefont {Q.}~\bibnamefont {Su}},\ and\
  \bibinfo {author} {\bibfnamefont {R.}~\bibnamefont {Grobe}},\ }\bibfield
  {title} {\bibinfo {title} {{N}umerical approach to solve the time-dependent
  {D}irac equation},\ }\href {https://doi.org/10.1103/PhysRevA.59.604}
  {\bibfield  {journal} {\bibinfo  {journal} {Phys. Rev. A}\ }\textbf {\bibinfo
  {volume} {59}},\ \bibinfo {pages} {604} (\bibinfo {year} {1999})}\BibitemShut
  {NoStop}%
\bibitem [{\citenamefont {Bauke}\ and\ \citenamefont
  {Keitel}(2011)}]{bauke_2011_GPU_acceleration_FFT_split_operator}%
  \BibitemOpen
  \bibfield  {author} {\bibinfo {author} {\bibfnamefont {H.}~\bibnamefont
  {Bauke}}\ and\ \bibinfo {author} {\bibfnamefont {C.~H.}\ \bibnamefont
  {Keitel}},\ }\bibfield  {title} {\bibinfo {title} {{A}ccelerating the
  {F}ourier split operator method via graphics processing units},\ }\href
  {https://doi.org/10.1016/j.cpc.2011.07.003} {\bibfield  {journal} {\bibinfo
  {journal} {Comput. Phys. Commun.}\ }\textbf {\bibinfo {volume} {182}},\
  \bibinfo {pages} {2454} (\bibinfo {year} {2011})}\BibitemShut {NoStop}%
\bibitem [{\citenamefont {Beerwerth}\ and\ \citenamefont
  {Bauke}(2015)}]{Beerwerth_2015_Krylov_subspace_methods}%
  \BibitemOpen
  \bibfield  {author} {\bibinfo {author} {\bibfnamefont {R.}~\bibnamefont
  {Beerwerth}}\ and\ \bibinfo {author} {\bibfnamefont {H.}~\bibnamefont
  {Bauke}},\ }\bibfield  {title} {\bibinfo {title} {{K}rylov subspace methods
  for the {D}irac equation},\ }\href
  {https://doi.org/10.1016/j.cpc.2014.11.008} {\bibfield  {journal} {\bibinfo
  {journal} {Computer Physics Communications}\ }\textbf {\bibinfo {volume}
  {188}},\ \bibinfo {pages} {189} (\bibinfo {year} {2015})}\BibitemShut
  {NoStop}%
\bibitem [{\citenamefont {Quesnel}\ and\ \citenamefont
  {Mora}(1998)}]{Quesnel_1998_gaussian_beam_coulomb_gauge}%
  \BibitemOpen
  \bibfield  {author} {\bibinfo {author} {\bibfnamefont {B.}~\bibnamefont
  {Quesnel}}\ and\ \bibinfo {author} {\bibfnamefont {P.}~\bibnamefont {Mora}},\
  }\bibfield  {title} {\bibinfo {title} {Theory and simulation of the
  interaction of ultraintense laser pulses with electrons in vacuum},\ }\href
  {https://doi.org/10.1103/PhysRevE.58.3719} {\bibfield  {journal} {\bibinfo
  {journal} {Phys. Rev. E}\ }\textbf {\bibinfo {volume} {58}},\ \bibinfo
  {pages} {3719} (\bibinfo {year} {1998})}\BibitemShut {NoStop}%
\bibitem [{Note1()}]{Note1}%
  \BibitemOpen
  \bibinfo {note} {Further details about adjusting the fields in \cite
  {Quesnel_1998_gaussian_beam_coulomb_gauge} can be found in the appendix of
  reference \cite {ahrens_guan_2022_beam_focus_longitudinal}.}\BibitemShut
  {Stop}%
\bibitem [{\citenamefont {Ahrens}(2012)}]{ahrens_2012_phdthesis_KDE}%
  \BibitemOpen
  \bibfield  {author} {\bibinfo {author} {\bibfnamefont {S.}~\bibnamefont
  {Ahrens}},\ }\emph {\bibinfo {title} {Investigation of the {K}apitza-{D}irac
  effect in the relativistic regime}},\ \href@noop {} {Ph.D. thesis},\ \bibinfo
   {school} {Ruprecht-Karls University Heidelberg} (\bibinfo {year} {2012}),\
  \bibinfo {note}
  {\url{http://archiv.ub.uni-heidelberg.de/volltextserver/14049/}}\BibitemShut
  {NoStop}%
\bibitem [{\citenamefont {Ahrens}\ and\ \citenamefont
  {Sun}(2017)}]{ahrens_2017_spin_non_conservation}%
  \BibitemOpen
  \bibfield  {author} {\bibinfo {author} {\bibfnamefont {S.}~\bibnamefont
  {Ahrens}}\ and\ \bibinfo {author} {\bibfnamefont {C.-P.}\ \bibnamefont
  {Sun}},\ }\bibfield  {title} {\bibinfo {title} {{S}pin in {C}ompton
  scattering with pronounced polarization dynamics},\ }\href
  {https://doi.org/10.1103/PhysRevA.96.063407} {\bibfield  {journal} {\bibinfo
  {journal} {Phys. Rev. A}\ }\textbf {\bibinfo {volume} {96}},\ \bibinfo
  {pages} {063407} (\bibinfo {year} {2017})}\BibitemShut {NoStop}%
\end{thebibliography}%

\end{document}